\documentclass[journal]{IEEEtran}

\usepackage{cite,graphicx,amsmath,amssymb}
\usepackage{subfigure}
\usepackage{citesort}
\usepackage{fancyhdr}
\usepackage{mdwmath}
\usepackage{mdwtab}
\usepackage{balance}
\usepackage{multirow}
\usepackage{eurosym}
\usepackage{xcolor}

\newtheorem{theorem}{Theorem}

\begin{document}

\title{User Association in 5G Networks: A Survey and an Outlook}

\author{Dantong\ Liu,~\IEEEmembership{Student Member,~IEEE,}  Lifeng\ Wang,~\IEEEmembership{Member,~IEEE,} Yue\ Chen,~\IEEEmembership{Senior Member,~IEEE,} Maged\ Elkashlan,~\IEEEmembership{Member,~IEEE,} Kai-Kit\ Wong,~\IEEEmembership{Fellow,~IEEE,} Robert\ Schober,~\IEEEmembership{Fellow,~IEEE,} and Lajos\ Hanzo,~\IEEEmembership{Fellow,~IEEE}
\thanks{D. Liu, Y. Chen, and M. Elkashlan are with the School of Electronic Engineering and Computer Science, Queen Mary University of London, London, E1 4NS, UK (email:\{d.liu, yue.chen, maged.elkashlan\}@qmul.ac.uk);}
\thanks{L. Wang and K. K. Wong are with the Department of Electronic \& Electrical Engineering, University College London, UK. (email: \{lifeng.wang, kai-kit.wong\}@ucl.ac.uk);}
\thanks{R. Schober is with the Institute for Digital Communications (IDC), Friedrich-Alexander-University Erlangen-Nurnberg ¡§
(FAU), Germany (email: schober@lnt.de);}
\thanks{L. Hanzo is with the School of Electronics and Computer Science, University of Southampton, Southampton, SO17 1BJ, UK (e-mail: lh@ecs.soton.ac.uk).}
}


\maketitle

\begin{abstract}
The fifth generation (5G) mobile networks are envisioned to support the deluge of data traffic with reduced energy consumption and improved quality of service (QoS) provision. To this end, key enabling technologies, such as heterogeneous networks (HetNets), massive multiple-input multiple-output (MIMO), and millimeter wave (mmWave) techniques, have been identified to bring 5G to fruition. Regardless of the technology adopted, a user association mechanism is needed to determine whether a user is associated with a particular base station (BS) before data transmission commences. User association plays a pivotal role in enhancing the load balancing, the spectrum efficiency, and the energy efficiency of networks. The emerging 5G networks introduce numerous challenges and opportunities for the design of sophisticated user association mechanisms. Hence, substantial research efforts are dedicated to the issues of user association in HetNets, massive MIMO networks, mmWave networks, and energy harvesting networks. We introduce a taxonomy as a framework for systematically studying the existing user association algorithms. Based on the proposed taxonomy, we then proceed to present an extensive overview of the state-of-the-art in user association algorithms conceived for HetNets, massive MIMO, mmWave, and energy harvesting networks. Finally, we summarize the challenges as well as opportunities of user association in 5G and provide design guidelines and potential solutions for sophisticated user association mechanisms.

\end{abstract}


\IEEEpeerreviewmaketitle

\begin{table}[!t]
\newcommand{\tabincell}[2]{\begin{tabular}{@{}#1@{}}#2\end{tabular}}
\renewcommand{\multirowsetup}{\centering}
\scriptsize
 \vspace{-0.2cm}
\centering
\caption{Summary of Abbreviation}
 \vspace{-0.2cm}
\begin{tabular}{|l|l|} \hline
5G & Fifth Generation \\
5G-PPP & 5G-Public Private Partnership\\
AP & Access Point\\
BBU & Base Band Unit\\
BPP & Binomial Point Process\\
BS & Base Station\\
CA & Carrier Aggregation \\
CAPEX & Capital Expenditure\\
CCO & Capacity and Coverage Optimization\\
CDF & Cumulative Distribution Function \\
CIR & Channel Impulse Response\\
CoMP & Coordinated Multipoint  \\
C-RAN &Cloud Radio Access Network\\
D2D & Device-to-Device\\
DL & Downlink\\
eICIC & enhanced Inter-Cell Interference Coordination\\
EU & European Union\\
FDD & Frequency-Division Duplex\\
HetNets & Heterogeneous Networks\\
ICIC & Inter-Cell Interference Coordination\\
LOS & Line-Of-Sight\\
LTE & Long Term Evolution\\
LTE-A & LTE-Advanced\\
LZFBF & Linear ZF Beamforming\\
M2M & Machine-to-Machine \\
MIMO & Multiple-Input Mutiple-Output\\
MLB & Mobility Load Balancing\\
MRT & Maximum Ratio Transmission \\
NLOS & Non-Line-Of-Sight \\
mmWave & millimeter Wave\\
OFDMA & Orthogonal Frequency-Division Multiple Access \\
OPEX & Operational Expenditure\\
PCP & Poisson Cluster Process\\
PPP & Poisson Point Process\\
QoS & Quality of Service\\
RF & Radio Frequency \\
RRH & Remote Radio Head\\
RSS & Received Signal Strength\\
SIR & Signal-to-Interference Ratio\\
SI & Self-Interference\\
SINR & Signal-to-Interference-plus-Noise Ratio\\
SNR & Signal-to-noise Ratio\\
SON & Self-Organizing Network\\
TDD& Time-Division Duplex \\
TPC & Transmit Pre-Coding \\
UA & User Association \\
UL & Uplink\\
WLAN & Wireless Local Area Network \\
WPAN & Wireless Personal Network\\
WPT & Wireless Power Transfer \\
ZF & Zero-Forcing\\\hline
\end{tabular}
\label{table:abb}
\vspace{-0.2cm}
\end{table}

\section{Introduction}
The proliferation of multimedia infotainment applications and high-end devices (e.g., smartphones, tablets, wearable devices, laptops, machine-to-machine communication devices) exacerbates the demand for high data rate services. According to the latest visual network index (VNI) report from Cisco~\cite{ciscowp}, the global mobile data traffic will increase nearly tenfold between 2014 and 2019, reaching 24.3 exabytes per month by 2019, wherein three-fourths will be video. Researchers in the field of communications have reached a consensus that incremental improvements fail to meet the escalating data demands of the foreseeable future. A paradigm shift is required for the emerging fifth generation (5G) mobile networks~\cite{AJG14}.

The intensifying 5G debate fuels a worldwide competition. To secure Europe's global competitiveness, in the seventh framework programme (FP7), the European commission has launched more than 10 European Union (EU) projects, such as METIS~\cite{METIS}, 5GNOW~\cite{5gnow}, iJOIN~\cite{iJOIN}, TROPIC~\cite{Tropic}, MCN~\cite{MCN}, COMBO~\cite{COMBO}, MOTO~\cite{MOTO}, and PHYLAWS~\cite{PHYLAWS} to address the architectural and functionality needs of 5G networks. Over the period from 2007 to 2013, the EU's investment into research on future networks amounted to more than {\euro}700 million, half of which was allocated to wireless technologies, contributing to the development of fourth generation (4G) and 5G systems. From 2014 to 2020, Horizon 2020~\cite{horizon}, which is the most extensive EU research and innovation programme, provides funding for the 5G-Public Private Partnership (5G-PPP)~\cite{5gppp}. To elaborate, 5G-PPP focuses on improving the communications infrastructure with an EU budget in excess of {\euro}700 million for research, development, and innovation over the next seven years. On the other hand, the governments of China and South Korea have been particularly devoted to pursuing 5G development efforts. In China, three ministries---the Ministry of Industry and Information Technology (MIIT), the National Development and Reform Commission (NDRC) as well as the Ministry of Science and Technology (MOST)---set up an IMT-2020 (5G) Promotion Group in February 2013 to promote 5G technology research in China and to facilitate international communication and cooperation~\cite{IMT}. Meanwhile, South Korea has established a 5G forum~\cite{5gforum}, which is similar to the EU's 5G-PPP, and committed \$1.5 billion for 5G development. Japan and the United States (US) have been less aggressive than the EU, China, and South Korea in setting national 5G research and development (R\&D) initiatives; nevertheless, the Japanese and US companies have also been proactive, with both academic institutions and enterprises taking up the 5G mantle. In May 2014, NTT DoCoMo announced plans to conduct ``experimental trials" of emerging 5G technologies together with six vendors: Alcatel-Lucent, Ericsson, Fujitsu, NEC, Nokia, and Samsung. The NYU wireless program~\cite{NYU}, launched in August 2012 by New York University's Polytechnic School of Engineering, is working on millimeter wave (mmWave) technologies and other research deemed crucial for 5G.

\begin{figure*} [tbp]
\vspace{-0.cm}
    \centerline{\includegraphics[width=0.9\textwidth,height=0.4\textwidth]{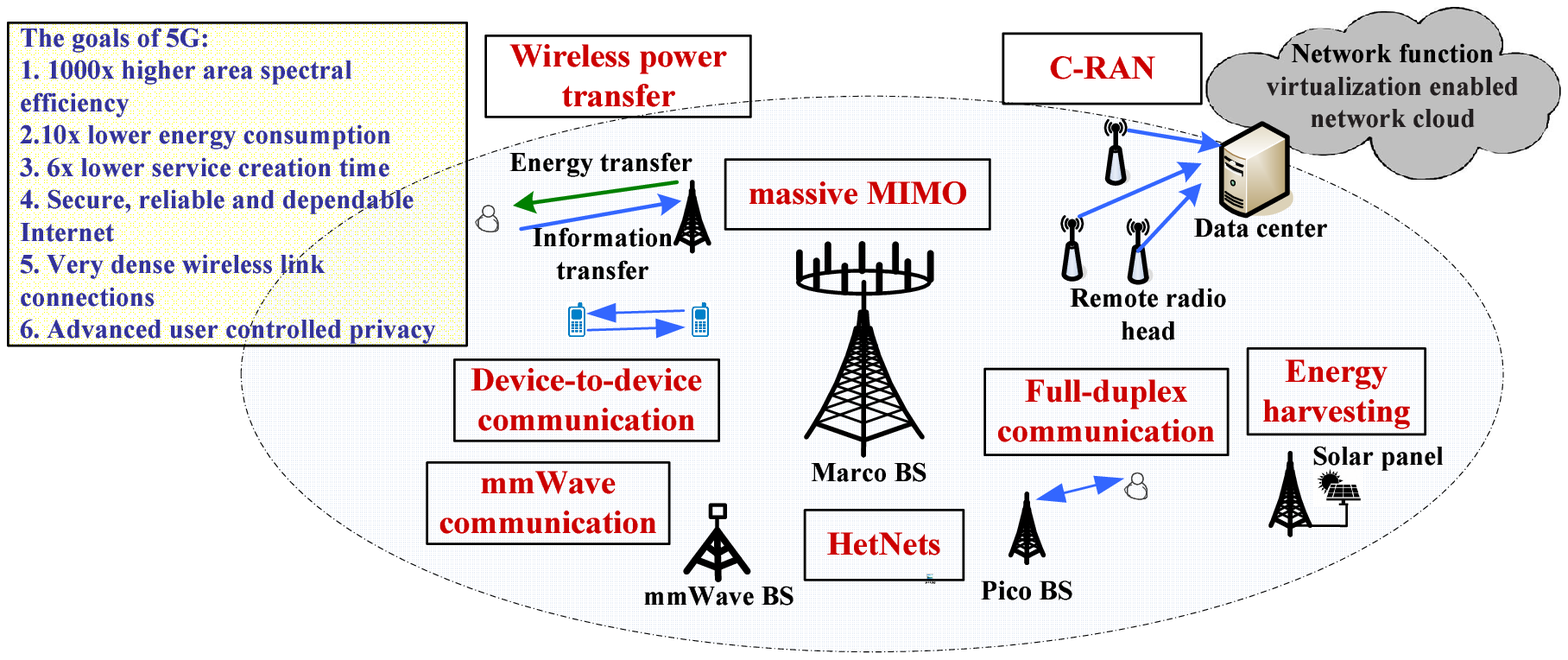}}
    \caption{Enabling technologies and expected goals of 5G networks.}
    \label{fig:5g}
\end{figure*}

The goals of 5G are broad, but are presumed to include the provision of at least 1,000 times higher wireless area spectral efficiency than current mobile networks. Other high-level key performance indicators (KPIs) envisioned by 5G-PPP include 10 times lower energy consumption per service, reduction of the average service creation time cycle from 90 hours to 90 minutes, creation of a secure, reliable, and dependable Internet with a ``zero perceived'' downtime for service provision, facilitation of very dense deployment of wireless communication links to connect over 7 trillion wireless devices serving over 7 billion people and enabling advanced user controlled privacy~\cite{5gppp}. To achieve these KPIs, the primary technologies and approaches identified by Hossain \emph{et al}.~\cite{HE15} for 5G networks are dense heterogeneous networks (HetNets), device-to-device (D2D) communication, full-duplex communication, massive multiple-input multiple-output (MIMO) as well as mmWave communication technologies, energy-aware communication and energy harvesting, cloud-based radio access networks (C-RANs), and the virtualization of wireless resources. More specifically, Andrews \emph{et al.}~\cite{AJG14} spotlight dense HetNets, mmWave, and massive MIMO as the ``big three'' of 5G technologies. Fig.~\ref{fig:5g} illustrates the enabling technologies and expected goals of 5G networks.

User association, namely associating a user with a particular serving base station (BS), substantially affects the network performance. {In the existing Long Term Evolution (LTE)/LTE-Advanced (LTE-A) systems, the radio admission control entity is located in the radio resource control layer of the protocol stack}, which decides whether a new radio-bearer admission request is admitted or rejected. The decision is made according to the quality of service (QoS) requirements of the requesting radio bearer, to the priority level of the request and to the availability of radio resources, with the goal of maximizing the radio resource exploitation~\cite{3GPP_TS36_300}. In existing systems, the received power based user association rule is the most prevalent one~\cite{3gppeicic10}, where a user will choose to associate with the specific BS, which provides the maximum received signal strength (max-RSS).  {The aforementioned new technologies and targets of 5G networks inevitably render such a rudimentary user association rule ineffective, and more sophisticated user association algorithms are needed for addressing the unique features of the emerging 5G networks. }

Numerous excellent contributions have surveyed the radio resource allocation issues in wireless networks. Explicit insights for understanding HetNets have been provided in~\cite{GA12,AJG13,AJ14}. A range of theoretical models, practical constraints, and challenges of HetNets were discussed in~\cite{GA12}. Seven key factors of the cellular paradigm shift to HetNets were identified in~\cite{AJG13}. An overview of load balancing in HetNets was given in~\cite{AJ14}. Additionally, the authors of~\cite{PM15} presented recent advances in interference control, resource allocation, and self-organization in underlay based HetNets. Lee~\emph{et al.}~\cite{YLL14} provided a comprehensive survey of radio resource allocation schemes for LTE/LTE-A HetNets. With the emphasis on green communication, Peng~\emph{et al.}~\cite{DF13} reviewed the emerging technologies conceived for improving the energy efficiency of wireless communications. The recent findings in energy efficient resource management designed for multicell cellular networks were surveyed in~\cite{RJB14}, while those designed for wireless local area networks (WLANs) and cellular networks were summarized in~\cite{BL14}. The benefits of BS sleep mode were considered in the same context in~\cite{WJ15}. In terms of network selection and network modeling,~\cite{LW13} studied the mathematical modeling of network selection in heterogeneous wireless networks relying on different radio access technologies. A suite of game-theoretic approaches developed for network selection were investigated in~\cite{TR12}. In~\cite{EH13}, stochastic geometry based models were surveyed in the context of both single-tier as well as multi-tier and cognitive cellular wireless networks.

While the aforementioned significant contributions have laid a solid foundation for the understanding of the diverse aspects of radio resource allocation in wireless networks, the resource allocation philosophy of 5G networks is far from being well understood. The following treatises have surveyed the recent advances in 5G networks, such as the key enabling technologies and potential challenges~\cite{HE15,BF14}, the architecture visions~\cite{AP14,CXW14}, the interference management~\cite{HE14}, and the energy efficient resource allocation~\cite{GW15}. Nevertheless, user association in 5G networks was not highlighted in these works. To highlight the significance of user association in 5G networks, this paper commences with a survey of user association in the context of the ``big three'' 5G technologies, defined in~\cite{AJG14}: HetNets, mmWave, and massive MIMO. We then address the important issues surrounding user association in energy harvesting networks. The challenges regarding user association in networks employing other 5G candidate technologies are also briefly elaborated on. Thereby, the contributions of this survey are fourfold, as summarized below.
\begin{enumerate}
  \item We present a comprehensive survey on the recent advances in user association algorithms designed for HetNets. The challenges imposed by the inherent nature of HetNets are identified. Many of those were largely ignored by most of the existing user association algorithms, although they have a great impact on the network performance.
  \item We investigate user association in the context of massive MIMO networks. The effects of massive MIMO on the received power, throughput, and energy efficiency are examined, and the existing solutions are reviewed. We highlight that the specific implementation of massive MIMO has a strong impact on user association and point out the fundamental aspects that should be carefully considered, when designing user association algorithms for massive MIMO networks.
  \item We study user association in mmWave networks. We highlight that the mmWave channel characteristics play a key role in the user association design. A range of important factors are illustrated in order to underline opportunities and challenges of this new field.
  \item We treat user association in energy harvesting networks, where we consider two specific energy harvesting mechanisms: energy harvesting from renewable energy sources and radio frequency (RF) energy harvesting. We also identify practical open challenges of user association in energy harvesting networks.
\end{enumerate}

In a dynamic scenario, a problem closely related to user association is the \emph{re-association/handover} problem. Deciding on  when to trigger a re-association/handover is an equally important problem, and understandably has gained significant attention~\cite{HZCJ15,HSXF14,HG2015}. In this survey, we focus our attention on the extensive review of user association in 5G networks, but disperse with the survey of re-association/handover.

\begin{figure} [tbp]
\vspace{-0.cm}
    \centerline{\includegraphics[width=0.5\textwidth,angle=-90]{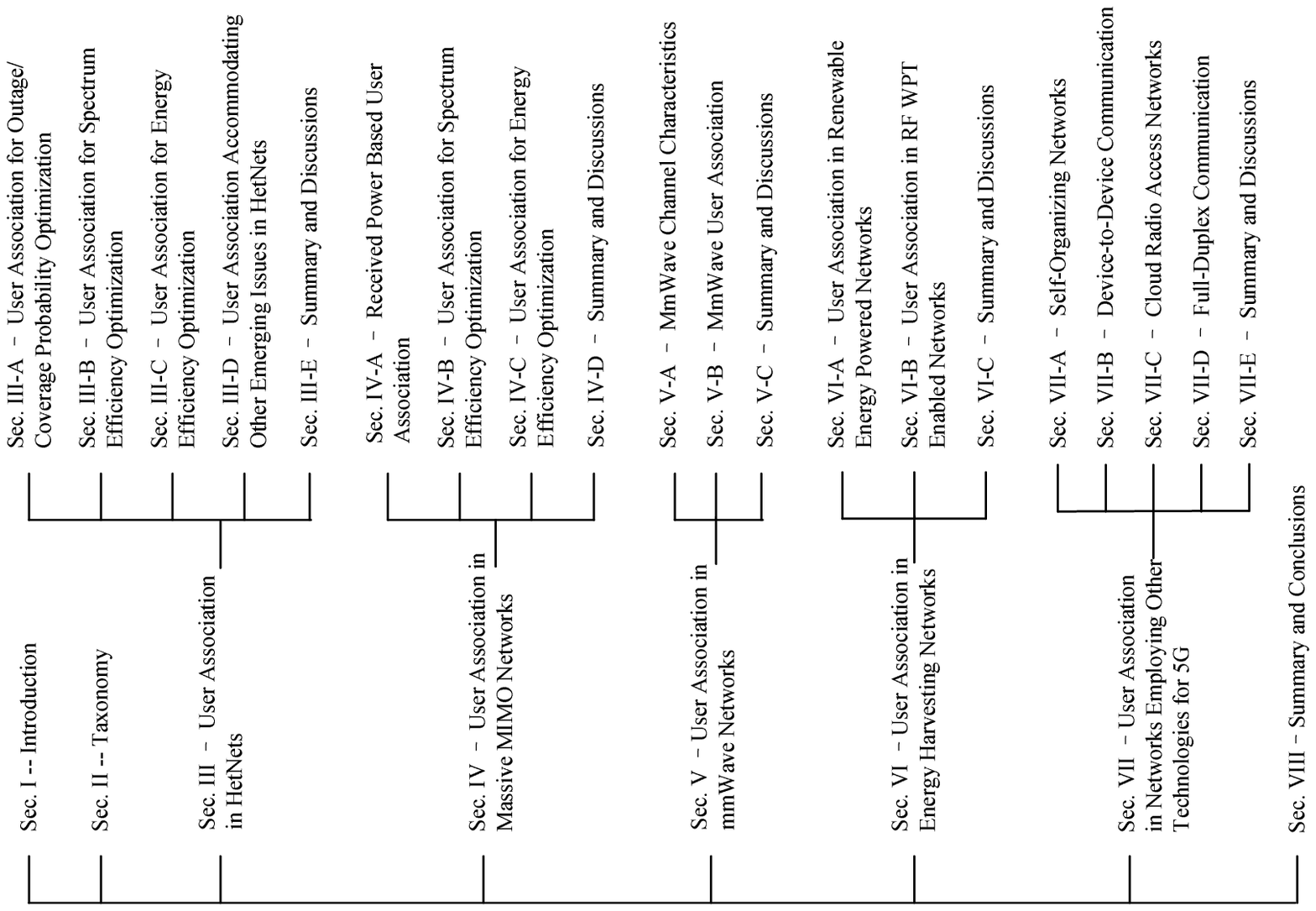}}
    \caption{The organization of this paper.}
    \label{fig:org}
\end{figure}

The rest of this paper is organized as follows. Section II presents a taxonomy which serves as a framework for systematically surveying the existing user association algorithms conceived for 5G networks. Section III elaborates on the existing user association algorithms for HetNets. The recent advances and open challenges of user association in massive MIMO and mmWave networks are discussed in Sections IV and V, respectively. In Section VI, user association algorithms designed for networks, which harvest energy from renewable energy sources and employ wireless power transfer (WPT) are surveyed. Section VII investigates the user association in networks employing other potential 5G technologies. Finally, the paper is concluded with design guidelines and conclusions in Section VIII and Section IX, respectively. For the sake of clarity, the organization of this paper is shown in Fig.~\ref{fig:org}

\section{Taxonomy}
\begin{figure*} [tb]
\vspace{-0.cm}
    \centerline{\includegraphics[width=0.85\textwidth]{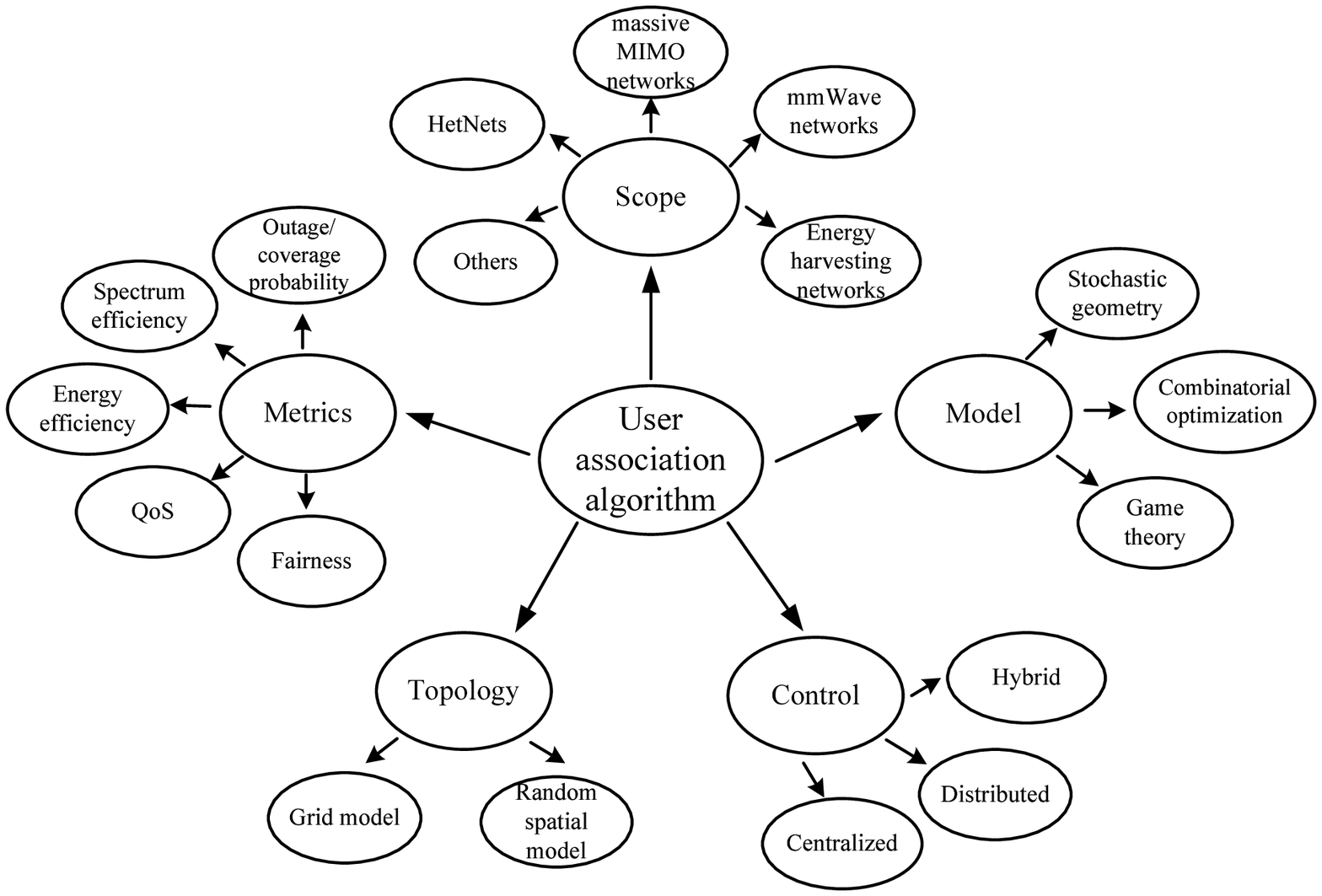}}
    \caption{The advocated taxonomy structure.}
    \label{fig:cato}
\end{figure*}

To systematically survey the existing user association algorithms for 5G networks, we develop a taxonomy, which could serve as a framework that accounts for all design aspects and may be used for evaluating the advantages and disadvantages of the proposed algorithms. Fig.~\ref{fig:cato} illustrates the advocated taxonomy, which consists of five non-overlapping branches: (1) Scope, (2) Metrics, (3) Topology, (4) Control, (5) Model. Each branch is further subdivided into different categories, as detailed below.

\subsection{Scope}
\begin{itemize}
\item
{\textbf{HetNets}\\
Cell densification constitutes a straightforward and effective approach of increasing the network capacity, which relies on densely reusing the spectrum across a geographical area and hence brings BSs closer to users. Specifically, the LTE-A standardization has already envisaged a multi-tier HetNets roll-out, which involved small cells underlaying macro-cellular networks. Small cells, such as picocells, femtocells and relays, transmit at a low power and serve as the fundamental element for the traffic offloading from macrocells, thereby improving the coverage quality and enhancing the cell-edge users' performance, whilst boosting both the area spectral efficiency and the energy efficiency~\cite{LP11}. As another benefit, this new palette of low power ``miniature'' BSs in small cells requires less upfront planning and lease costs, consequently reducing both the network's operational and capital expenditures (OPEX, CAPEX)~\cite{CV08}. The following details the features of small cells in HetNets.
\begin{itemize}
  \item {Picocells are covered by low-power operator-installed BSs relying on the same backhaul and access features as macrocells. They are usually deployed in a centralized manner, serving a few tens of users within a radio range of say 300 m or less, and have a typical transmit power ranging from 23 to 30 dBm~\cite{LP11}. Picocells do not require an air conditioning unit for the power amplifier, and incur much lower cost than traditional macro BSs~\cite{DA11}.}
  \item {Femtocells, also known as home BSs or home eNBs, are low-cost, low-power, and user-deployed access points. Typically, a femtocell's coverage range is less than 50 m and its transmit power is less than 23 dBm. Femtocells operate in open or restricted (closed subscriber group) access~\cite{LP11}.}
  \item Relays are usually operator-deployed access points that route data from the macro BS to users and vice versa~\cite{LP11}. Relays are typically connected to the rest of the network via a wireless backhaul. They can be deployed both indoors and outdoors, with the transmit power ranging from 23 to 33 dBm for outdoor deployment, and 20 dBm or less for indoor deployment~\cite{DA11}.
\end{itemize}}
\item
{\textbf{Massive MIMO Networks}\\
Conventional MIMO is unable to achieve the high multiplexing gains required to meet the 5G KPIs, due to the limited number of antennas. By contrast, massive MIMO BSs with large antenna arrays are potentially capable of serving dozens of single-antenna users over the same time and frequency range~\cite{Emil_BJ_10myths_2015}. The main features of massive MIMO are as follows.
  \begin{itemize}
    \item Massive MIMO achieves a high power-gain, hence significantly increasing the received signal power. Therefore, it necessitates a reduced transmit power to achieve a required QoS~\cite{ngo2013energy}.

    \item Massive MIMO exhibits a high spectrum efficiency, which substantially improves the throughput. This is attributed to the fact that BSs having large antenna arrays are capable of serving more users~\cite{EG_Larsson_COMMAG_2014}.

    \item {Channel estimation errors, hardware impairments, and small-scale fading effects are averaged out when the number of BS antennas is sufficiently high~\cite{Hoydis2013}.} However, the so-called pilot contamination becomes the main performance limitation, which is due to reusing the same pilot signals in adjacent cells~\cite{Hoydis2013}.
  \end{itemize}
}
\item
{\textbf{MmWave Networks}\\
Due to its large bandwidth, mmWave supports Gigabit wireless services. As mentioned in \cite{S_Singh2014}, mmWave can be a scalable solution for future wireless backhaul networks. MmWave transmission has been adopted in several standards such as IEEE 802.15.3c~\cite{IEEE_802_15_3c} for indoor wireless personal networks (WPANs) and IEEE 802.11ad~\cite{802_11ad_st} for WLANs. As one of the key 5G techniques, mmWave systems exhibit the following features.
  \begin{itemize}
   \item  Compared to the traditional low frequency communication systems, the path-loss experienced by high-frequency mmWave signals is increased by several orders of magnitude~\cite{Ayach2014}. Hence, mmWave transmission is only suitable for short-range systems.
   \item In mmWave systems, highly directional communication relying on narrow beams is employed for achieving a high beamforming gain and for suppressing the interference arriving from neighboring cells~\cite{TSR2013}.
   \item For a fixed array aperture, mmWave BSs pack more antennas into a given space and hence attain an increased array gain. They also adopt low-complexity analog beamforming/precoding schemes due to hardware constraints experienced at these high frequencies~\cite{ZP2011}.
 \end{itemize}}

 \item
{\textbf{Energy Harvesting Networks}\\
Recent developments in energy harvesting technologies have made the dream of self-sustaining devices and BSs potentially possible. As such, energy harvesting is highly desirable both for prolonging the battery life and for improving the energy efficiency of networks~\cite{HE15,DZ15}. According to the specific sources of the harvested energy, energy harvesting networks may be categorized as follows.
\begin{itemize}
   \item BSs and users may harvest renewable energy from the environment, such as solar energy or wind energy~\cite{HE15,LH13}. However, the renewable energy is volatile, e.g., the daily solar energy generation peaks around noon, and decays during the later part of the day. This inherent intermittent nature of renewable energy challenges the reliable QoS provision in wireless networks~\cite{DL14PIMRC,TH13,DLWCNC}.
  \item Alternatively, BSs and users can harvest energy from ambient radio signals, relying on RF energy harvesting~\cite{CheDZ14,Yuanwei2014_globcom,huangkaibin_mag_2015,S_A_H_WPT_2015,H_T_WPT_2015,Lifeng_wang2015_globcom}. In this context, simultaneous wireless information and power transfer is envisaged as a promising technology for 5G wireless networks~\cite{HE15}.
 \end{itemize}}

  \item
\textbf{{Other 5G Candidate Technologies}}\\
{Apart from the aforementioned network types, networks employing other candidate technologies may also constitute imperative part of the wireless evolution to 5G. They are elaborated as follows.
\begin{itemize}
  \item Self-Organizing Networks (SONs) have the ability of self-configuration, self-optimization, and self-healing, which minimize the level of manual work. For these networks, multiple use cases are identified for network optimization~~\cite{Fehske_AJ_2013}.
  \item Device-to-Device (D2D) communication allows direct transmission between devices for improving the spectrum efficiency and energy efficiency. One of the key features in D2D communication is that it is controlled by BSs~\cite{Asadi_A_2014}.
  \item Cloud Radio Access Networks (C-RAN) move the baseband units to the cloud for centralized processing, which significantly reduces CAPEX and OPEX. In the C-RAN,  the backhaul between remote radio heads (RRHs) and base band units (BBUs) forms a key component~\cite{checko_2015}.
  \item Full-duplex communication supports the downlink and uplink transmission at the same time and frequency resource, which enhances the spectrum efficiency. However, self-interference suppression plays a key role in full-duplex communication~\cite{Zhongshan_2015}.
 \end{itemize}}
\end{itemize}

\subsection{Metrics}
For user association in 5G networks, different metrics have been adopted for determining which specific BS should serve which user. Five metrics are commonly used in this context: Outage/coverage probability, spectrum efficiency, energy efficiency, QoS, and fairness. In the related research literature, either one of the specified metrics or a combination of several of them is used.
\begin{itemize}
\item
{\textbf{Outage/coverage probability}: A crucial aspect in the evaluation and planning of a wireless network is the effect of co-channel interference imposed on radio links. The probabilities that the signal-to-interference-plus-noise ratio (SINR) drops below and rises above a certain threshold are defined as outage probability and coverage probability, respectively. The outage/coverage probability is crucial in terms of benchmarking the average throughput of a randomly chosen user in the network, and serves as a fundamental metric for network performance analysis and optimization~\cite{EH13,DHS12,DHS13}.}

\item
{\textbf{Spectrum efficiency}: Spectrum efficiency refers to the maximum information rate that can be transmitted over a given bandwidth in a specific communication system. With the surge of data traffic and limited spectrum resources, a high spectrum efficiency is a mandatory requirement of 5G networks~\cite{HRQ14}.}

\item
{\textbf{Energy efficiency}: Driven by environmental concerns, green communication has drawn tremendous attention from both industry and academia~\cite{HRQ14,BL14}. Various energy efficiency metrics have been adopted in the literature to provide a quantitative analysis of the power saving potential of a certain algorithm. There are two main types of energy efficiency metrics:
\begin{enumerate}
  \item {The ratio between the total data rate of all users and the total energy consumption (bits/Joule)~\cite{MA14ICC,LS13,DL14VTC}.}
  \item {The direct presentation of the power/energy saving achieved by means of a certain algorithm (e.g., the difference in power/energy consumption before and after the adoption of a certain algorithm, the percentage of power saving, etc.)~\cite{HZ12,LS13,CRE15}.}
\end{enumerate}}
\item
{\textbf{QoS}: {As the salient performance metric experienced by users of the network, the QoS is  of primary concern for network operators, whilst maintaining profitability. The QoS provision can be quantitatively measured in terms of the traffic delay~\cite{DL14PIMRC}, the user throughput~\cite{DL14CCNC,MA14ICC,DL14letter}, the SINR~\cite{YSS13,WCC12}, etc., in order to cater for the heterogeneous requirements of today's and tomorrow's diverse multimedia infotainment applications and broadband-hungry mobile devices.}
}
\item
{\textbf{Fairness}: Facilitating fairness amongst users constitutes another important issue in the radio resource allocation of wireless networks. The traditional fairness problem is related to packet scheduling among users, where each user should receive a fair amount of radio resources for his/her wireless access. In HetNets, the fairness problem arises not only in scheduling within a traditional cell but also in the user association decision among cells in different tiers. Specifically, if radio resources are allocated on the basis that the lowest achievable rate among users is maximized, the allocation is said to be max-min fair~\cite{WCL14,DBLP2014}. In other words, users with a poor channel quality will receive more radio resources and those having a good channel quality will receive a smaller proportion of radio resources. To evaluate fairness, the Jain's fairness index~\cite{RJ91} has been widely adopted~\cite{DL14CCNC,DL14letter}, which is defined as
\begin{align}\label{FAIRNESS}
\mathcal{J}\left( {{r_1}, \cdots {r_n}, \cdots {r_N}} \right) = \frac{{{{\left( {\sum\nolimits_{n = 1}^N {{r_n}} } \right)}^2}}}{{N\sum\nolimits_{n = 1}^N {r_n^2} }}.
\end{align}
Jain's fairness index rates the fairness of a set of values where $N$ is the number of users and $r_n$ is the throughput of the $n$-th user.
}
\end{itemize}

\subsection{Topology}
\begin{figure}[t!]
    \begin{center}
        \includegraphics[width=3in,height=2in]{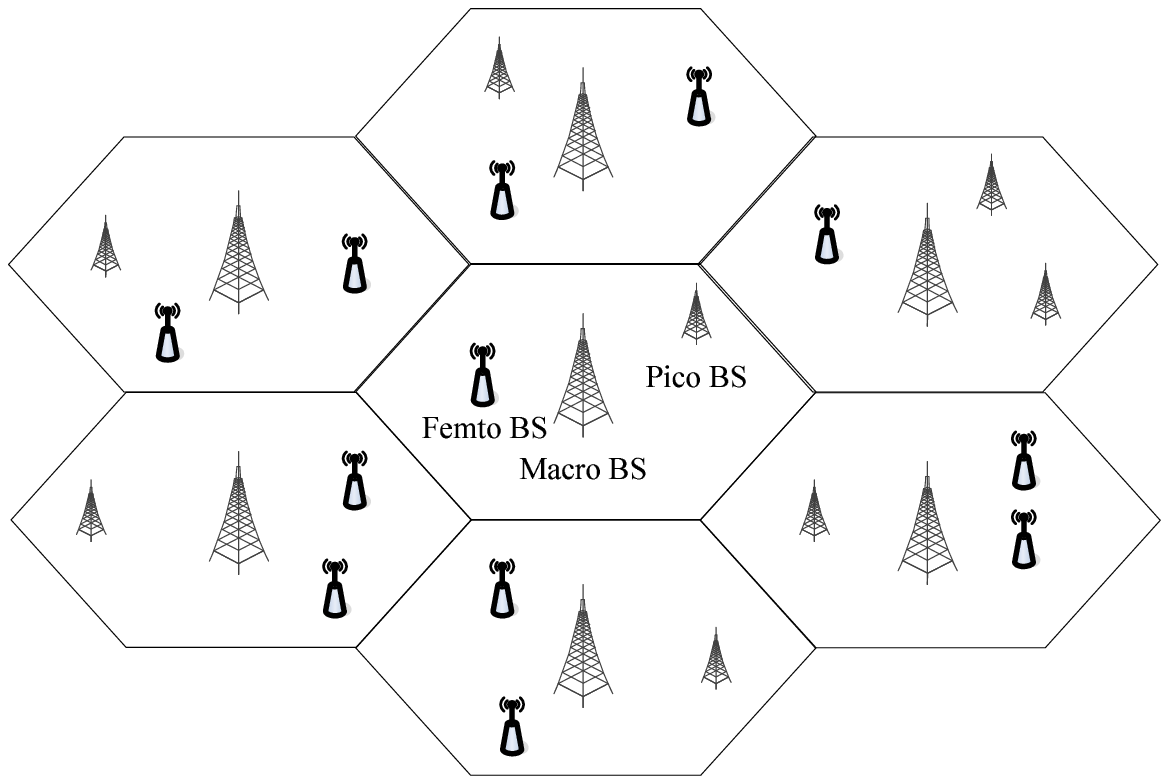}
        \caption{A 3-tier HetNet topology for the grid model, where the macro BSs are located in the center of the hexagonal cell with the pico and femto BSs located along the macro BSs.}
        \label{grid}
    \end{center}
\end{figure}

\begin{figure}[t!]
    \begin{center}
        \includegraphics[width=3in,height=2 in]{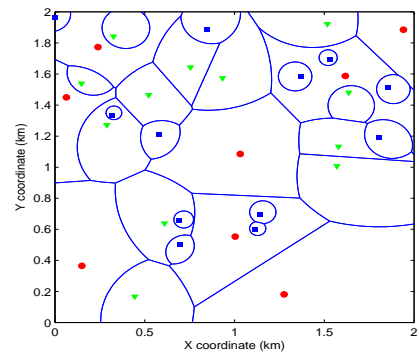}
        \caption{A 3-tier HetNet following the random spatial model from stochastic geometry, where the macro BSs (red circle) are overlaid with pico BSs (green triangle) and femto BSs (blue square).}
        \label{random}
    \end{center}
\end{figure}

Currently, there are two distinct approaches in modeling the topology of networks.
\begin{itemize}
\item
{\textbf{Grid model}: The grid model is widely used in the research of radio resource allocation in wireless networks, where all BSs are assumed to be located on a regular grid, (e.g., the traditional hexagonal grid model). For such a model, time-consuming Monte Carlo simulations are required for performance evaluation, and their mathematical analysis is often intractable.}
\item
{\textbf{Random spatial model}: The random spatial model is an emerging approach of modeling the topology of wireless networks. This model is capable of capturing the topological randomness in the network geometry. Employing tools from stochastic geometry, simple closed-form expressions can be derived for key performance metrics, which leads to tractable analytical results~\cite{EH13}. However, the accuracy of the results largely depends on whether the adoption of point processes routinely used in stochastic geometry is capable of appropriately capturing the characteristics of real network conditions.}
\end{itemize}
Fig.~\ref{grid} and Fig.~\ref{random} show a 3-tier HetNet topology for the grid model and the random spatial model, respectively.

\subsection{Control}
The adopted control mechanisms heavily affect the computational complexity, the signaling overhead, and the optimality of user association algorithms. Broadly speaking, there are three different control mechanisms.
\begin{itemize}
\item
{\textbf{Centralized control}: In the centralized approach, the network contains a single central entity that performs resource allocation. This central entity collects information, such as the channel quality and the resource demand from all users. Based on the information obtained, the central entity decides which particular BS is to serve which user~\cite{YRX14,FD11,GJ12}. Centralized control is capable of providing optimal resource allocation for the entire network and exhibits a fast convergence, but the required amount of signaling may be excessive for medium to large-sized networks.
\item
\textbf{Distributed control}: By definition, distributed control does not require a central entity and allows BSs and users to make autonomous user association decisions by themselves through the interaction between BSs and users. Hence, distributed control is attractive owing to its low implementational complexity and low signaling overhead. It is particularly suitable for large networks, especially for HetNets associated with many autonomous femtocells~\cite{QY12,KS14}. However, for distributed control, users or BSs make autonomous decisions in a distributed manner, which may lead to the ``Tragedy of the Commons''~\cite{GH68}. Explicitly, this describes a dilemma in which multiple individuals acting independently in their own self-interest, ultimately jam the limited shared resources even when it is clear that it is not in anyone's long term interest for this to happen.
\item
\textbf{Hybrid control}: Hybrid control relies on a compromise approach, which combines the advantages of both centralized and distributed control. For instance, the power control at the BS may rely on using a distributed method, whereas load balancing across the entire network could be implemented in a centralized manner~\cite{HZ12}.}
\end{itemize}

\subsection{Model}

Utility is widely employed for modeling the user association problem. For making a decision, utility quantifies the satisfaction that a specific service provides for the decision maker~\cite{PF70}. Depending on the metric adopted, the utility relied upon in user association may be constituted of spectrum efficiency, energy efficiency, QoS, etc. In some recent studies, logarithmic~\cite{QY12,DL14VTC}, exponential~\cite{YYJZ11,DL14letter}, and sigmoidal~\cite{LCBW11} utility functions are utilized to model these attributes. By contrast, for studies which do not specifically discuss the choice of utility functions, we may safely assume that they use linear utility functions, namely that the utility is the spectrum efficiency, energy efficiency, or QoS itself. Combined with the utility based design, game theory, combinatorial optimization, and stochastic geometry are the prevalent tools popularly adopted for solving the user association problem.

\begin{itemize}
\item
{\textbf{Game theory}: Game theory is a mathematical modeling tool, which has distinct advantages in investigating the interaction of multiple players. The combination of strategies incorporating the best strategy for every player is known as \emph{equilibrium}~\cite{LW13}. In particular, the solution of the game achieves \emph{Nash Equilibrium}, if none of the players can increase its utility by changing his or her strategy without degrading the utility of the others~\cite{TR12}. Hence, game theory is a powerful tool which is also capable of solving user association problems. In this context, the players can be the BSs as in~\cite{DL14CCNC,DL14letter,DL14VTC} or the users as in~\cite{VNH14} or both as in~\cite{SO14,NN14,HM13,SW14,MH13}, and the strategies are constituted of the corresponding user association decisions. In non-cooperative game theory modeling as in~\cite{SO14,NN14,PF13,VNH14,MH13}, the players seek to maximize their own utility and compete against each other by adopting different strategies, such as adjusting their transmit powers~\cite{VNH14} or placing bids representing the willingness to pay~\cite{MH13}. By contrast, cooperative game theory models a bargaining game as in~\cite{DL14CCNC,DL14letter,DL14VTC}, where the players bargain with each other for the sake of attaining mutual advantages. Game theory is suitable for designing distributed algorithms endowed with flexible self-configuration features, despite only imposing a low communication overhead~\cite{TR12}. However, it is important to note that game theory operates under the assumption of rationality, that is, all players are rational individuals acting in their own best interest. {However, in 5G networks, players --- BSs or users --- can not be guaranteed to act in a rational manner all the time~\cite{SLHT11}. For example, BSs involved in the game may have different optimization objectives, the one maximizing its energy efficiency will perhaps be perceived as non-rational by the other one maximizing its transmission rate and vice versa.}}
\item
{\textbf{Combinatorial optimization}:}
Utility maximization under resource constraints constitutes a general modeling approach for user association in 5G networks, which is formulated as:
\begin{align}\label{1}
\begin{array}{l}
\mathop {\max }\limits_{\bf{x}}\;\mathcal{U }= \sum\nolimits_{m = 1}^M {\sum\nolimits_{n = 1}^N {{x_{mn}}{\mu _{mn}}} } ,\\
{\rm{s}}{\rm{.t}}{\rm{.}}\quad {f_i}\left( {\bf{x}} \right) \le {c_i},\;i = 1, \cdots, p,
\end{array}
\end{align}
where ${\bf{x}} = [{x_{mn}}]$ is the user association matrix, in which ${x_{mn}}=1$, if user $n$ is associated with BS $m$, otherwise ${x_{mn}}=0$; $\mathcal{U}$ is the total network utility; ${\mu _{mn}}$ is the utility of user $n$, when associated with BS $m$; ${f_i}\left( {\bf{x}} \right) \le {c_i}$ represents the resource constraints, such as spectrum constraints, power constraints, QoS requirements, etc. Since we normally assume that a specific user can only be associated with a single BS at any time, that is ${x_{mn}} = \left\{ {0,1} \right\}$, the resultant problem is a combinatorial optimization problem, which is in general NP-hard. In other words, performing an exhaustive search for solving the problem optimally is computationally prohibitive even for medium-sized networks. A popular method of overcoming this issue is to make the problem convex by relaxing the user association matrix from ${x_{mn}} = \left\{ {0,1} \right\}$ to ${x_{mn}} = [0,1]$. Then, the classic Lagrangian dual analysis~\cite{convex} can be invoked, followed by recovering the primal user association matrix ${\bf{x}}$ from the optimal dual problem. However, due to the discrete nature of primal combinatorial optimization, the relaxation of the user association matrix ${\bf{x}}$ may lead to a duality gap between the primal and dual problems~\cite{QY12,KS14,Dantong_massivMIMO_2015}.

\item
\textbf{Stochastic geometry}: Stochastic geometry constitutes an emerging modeling approach, which not only captures the topological randomness of the network geometry, but serendipitously leads to tractable analytical results. Stochastic geometry is a powerful mathematical and statistical tool conceived for the modeling, analysis, and design of wireless networks relying on random topologies~\cite{EH13}. In stochastic geometry based analysis, the network is assumed to obey a certain point process, which captures the network properties. More explicitly, depending on the particular network type as well as on the MAC layer behavior, a closely matching point process is selected for modeling the positions of the network entities. Examples of particular point processes include the Poisson point process (PPP), the Binomial point process (BPP), the Hard core point process (HCPP), and the Poisson cluster process (PCP), whose detailed definition can be found in~\cite{PPP12,EH13}. Based on the specific properties of the selected point process, analytical expressions can be derived for the interference, for the coverage probability, for the outage probability, etc.~\cite{DHS12,DHS13,WCC12}. However, the performance metrics considered in most of the recent treatises are mainly based on Shannon's capacity formula. Despite the rich literature, the adoption of point processes that accurately capture the characteristics of 5G networks is still an open research challenge.

\end{itemize}

\section {User Association in HetNets}
Dense HetNets are likely to become the dominant theme during the wireless evolution towards 5G~\cite{BN14}. However, the conventional max-RSS user association rule is unsuitable for HetNets, since the transmit power disparity of marcocells and small cells will lead to the association of most of the users with the macro BS~\cite{3gpp10}, hence potentially resulting in inefficient small cell deployment.

To cope with this problem, the concept of biased user association has been proposed by 3GPP in Release 10~\cite{3gppCRE}, where the users' power received from the small cell BSs is artificially increased by adding a bias to it to ensure that more users will be associated with small cells. In~\cite{GI11}, the macro-to-small cell off-loading benefits of biased user association were demonstrated in terms of the attainable capacity improvement. However, the drawback of biased user association is that the group of users, who are forced to be associated with small cells owing to the added bias, experience strong interference from the nearby macrocell~\cite{HSJ11}. In this context, the improvement achieved by offloading traffic to small cells might be offset by the strong interference. Therefore, the trade-off between network load balancing and network throughput strictly depends on the value of the selected bias, which has to be carefully optimized in order to maximize the network utility~\cite{HE14}. {In~\cite{KUTO13}, Q-learning was used for determining the bias value of each user, where each user independently learns from past experience the bias value that minimizes the number of users in outage.} Moreover, several interference mitigation schemes based on resource partitioning have been proposed for solving the above problem in biased user association, including the inter-cell interference coordination (ICIC) technique proposed in 3GPP Release 8 and the enhanced inter-cell interference coordination (eICIC) solution advocated in 3GPP Release 10~\cite{3gppeicic}. The authors of \cite{DS14} optimized both the bias value and the resource partitioning in eICIC enabled HetNets.

In this section, the existing research results on user association in HetNets are surveyed and categorized according to a diverse range of different performance metrics, as summarized in Table~\ref{table:1}, which provides a qualitative comparison of all user association algorithms conceived for HetNets and discussed in this section. In Table~\ref{table:1}, ``-'' means that the corresponding algorithm did not consider this metric, ``UA'' stands for user association, ``DL'' and ``UL'' represent downlink and uplink, respectively. Additionally, a range of open challenges in user association for HetNets are highlighted.

\begin{table*}[tb]
\newcommand{\tabincell}[2]{\begin{tabular}{@{}#1@{}}#2\end{tabular}}
\renewcommand{\multirowsetup}{\centering}
\scriptsize
 \vspace{-0.2cm}
\centering
\caption{Qualitative Comparison of User Association Algorithms for HetNets}
 \vspace{-0.2cm}
\begin{tabular}{|l|r|r|r|r|r|r|r|r|r|r|} \hline
\textbf{Ref.}&\textbf{Algorithm}&\textbf{Topology} & \textbf{Model}&\textbf{Direction}&\textbf{Control}&\textbf{\tabincell{r}{Spectrum\\efficiency}} & \textbf{\tabincell{r}{Energy\\efficiency}}&\textbf{\tabincell{r}{QoS\\provision}} & \textbf{Fairness}&\textbf{\tabincell{r}{Coverage\\probability}}\\ \hline
\cite{DHS12}& max-RSS UA& \multirow{9}{0.9cm}{Random spatial}& \multirow{9}{1.1cm}{Stochastic geometry}&DL&Distributed&Low&-&Moderate&-&Low \\
\cite{DHS13}& max-RSS UA& & &DL&Distributed&Moderate&-&Moderate&-&Moderate \\
\cite{WCC12}& \tabincell{c}{max-RSS UA+\\spectrum partitioning}& &  &DL&Distributed&High&-&High&-&High\\
\cite{HS12}& biased UA& & &DL&Distributed&Moderate&-&Moderate&-&Moderate\\
\cite{SS13}& biased UA& & &DL&Distributed&Moderate&-&High&-&High\\
\cite{SSJG14}&{\tabincell{c}{biased UA+spectrum partitioning}}& & &DL&Distributed&High&-&Moderate&-&Moderate\\
\cite{WBBL14,SSRS14}&{\tabincell{c}{biased UA+spectrum partitioning}}& & &DL&Distributed&High&-&Moderate&-&High\\
\cite{LYBW15}&{\tabincell{c}{biased UA+spectrum partitioning}}& & &DL/UL&Distributed&High&-&Moderate&-&High\\
\cite{YSS13}& biased UA+BS sleeping& & & DL&Distributed&-&High&High&-&High\\ \hline

\cite{CS12}& UA & \multirow{12}{0.7cm}{Grid} & \multirow{12}{1.3cm}{Combinatorial optimization}& DL& Centralized& High& -& Moderate&Low&- \\
\cite{QY12}& UA &  & & DL& Distributed& Moderate& -& Moderate&High&- \\
\cite{HZSM15}& UA &  & & DL& Centralized& High& -& -&-&- \\
\cite{MA14ICC}&UA& & & DL&Distributed&-&High&High&Low&-\\
\cite{YRX14}&UA& & & DL&Centralized&High&High&High&-&-\\
\cite{FD11,GJ12}& UA+spectrum partitioning& &  &DL&Centralized&Moderate&-&Moderate&High&-\\
\cite{MJF14}& UA+spectrum partitioning& &  &DL&Centralized&High&-&Moderate&Moderate&-\\
\cite{KS14}& UA+power control&  & & DL& Distributed& High& -& Moderate&High&- \\
\cite{MR10,SR15}& UA+power control&  & & DL& Distributed& High& -& High&High&- \\
\cite{PH13}&UA+power control& & & UL&Centralized&-&High&High&Low&-\\
\cite{HZ12}&UA+power control& & & DL&Hybrid&-&High&High&Moderate&-\\
\cite{LPQ13}&UA+power control& & & DL&Centralized&Moderate&High&-&-&-\\
\cite{LS13} &UA+BS sleeping& & & DL&Centralized&-&High&Moderate&High&-\\
\cite{CRE15} &UA+BS sleeping& & & DL&Distributed&-&High&High&-&-\\ \hline
\cite{DL14CCNC}&UA &\multirow{7}{0.7cm}{Grid} & \multirow{7}{1.2cm}{Game theory}&DL& Distributed&Moderate& -& Moderate&High&- \\
\cite{DL14letter}&UA & & &DL& Distributed&Moderate& -& High&High&- \\
\cite{SO14}&UA & & &DL& Distributed&High& -& Moderate&High&- \\
\cite{NN14,PF13}&UA && &DL& Distributed&Moderate& -& High&Moderate&-\\
\cite{SW14}&UA && &UL& Distributed&Moderate& -& High&Moderate&-\\
\cite{VNH14}& UA+power control& & & UL&Distributed& High& -& High&High&- \\
\cite{MH13} &UA+power control& & & UL&Distributed& High& -& Moderate&High&-\\\hline
\end{tabular}
\label{table:1}
\vspace{-0.2cm}
\end{table*}

\subsection {User Association for Outage/Coverage Probability Optimization}
The outage/coverage probability is used for evaluating the performance of the desired user in wireless networks. In fact, the outage/coverage probability is the primary performance metric employed for user association analysis in conjunction with stochastic geometry. In particular, the authors of \cite{DHS12,DHS13} modeled and analyzed the performance of max-RSS user association in K-tier downlink HetNets with the aid of stochastic geometry. The coverage probability of interference limited underlay HetNets was presented, and the nature of cell loads experienced in K-tier HetNets was demonstrated in~\cite{DHS12}. The authors showed that due to the high load difference amongst the coexisting network elements, some network elements might be idle and hence would not contribute to the aggregate interference level. Therefore, the SINR model of \cite{DHS12} was improved in~\cite{DHS13} in order to account for the activity factor of the coexisting heterogeneous BSs. It was shown that adding lightly-loaded femtocells and picocells to the network increases the overall coverage probability. However, due to the random deployment of small cells coupled with the high transmit power difference with regard to the macro BSs, there might be some overloaded network elements (i.e., marcocell) and a large number of under-utilized small cells. By relying on an approach similar to the one used in~\cite{DHS12,DHS13}, the authors of \cite{WCC12} first derived the coverage probability for each tier under different spectrum allocation and femtocell access policies, and then formulated the throughput maximization problem subject to specific QoS constraints expressed in terms of both coverage probabilities and per-tier minimum rates. The results provided beneficial insights into the optimal spectrum allocation.

The effect of biased user association was investigated in the context of multi-tier downlink HetNets in~\cite{HS12} and~\cite{SS13} with the aid of stochastic geometry, where the optimal bias resulting in the highest signal-to-interference ratio (SIR) and the highest rate coverage were determined using numerical evaluation techniques. Biased user association and spectrum partitioning between the macrocell and small cells were considered in~\cite{SSJG14,WBBL14,SSRS14}. The authors of~\cite{SSJG14} analyzed the coverage guaranteeing a certain throughput for a two-tier topology and provided insights concerning the most appropriate spectrum partitioning ratio based on numerical investigations. For a general multi-tier network, spectrum partitioning and user association were optimized in the downlink analytically in terms of the attainable coverage probability in~\cite{WBBL14} and the coverage guaranteeing a certain throughput in~\cite{SSRS14}. In contrast to the aforementioned works on downlink HetNets, in~\cite{LYBW15} the optimal user association bias and spectrum partitioning ratios were derived analytically for the maximization of the proportionally fair utility of the network based on the coverage probability both in the downlink and uplink of HetNets. The results revealed that the optimal uplink and downlink user association biases are not identical, thereby reflecting the tradeoff between uplink and downlink performance in HetNets, when the users are constrained to associate with the same BS in both uplink and downlink.

A qualitative comparison of the above-mentioned user association algorithms for coverage probability optimization in HetNets is provided in Table~\ref{table:1}.

\subsection {User Association for Spectrum Efficiency Optimization}
Spectrum efficiency is a widely accepted network performance metric. In~\cite{CS12}, dynamic user association was proposed for the downlink of HetNets in order to maximize the sum rate of all users. The authors derived an upper bound on the downlink sum rate using convex optimization and then proposed a heuristic user association rule having a low complexity and approaching the performance upper bound. Their simulation results verified the superiority of the proposed heuristic user association rule over the classic max-RSS and biased user association in terms of the average user data rate. However, it is widely recognized that maximizing the sum data rate of all users may result in an unfair data rate allocation, which was also reflected by the results of~\cite[Fig. 3]{CS12}. Based on~\cite[Fig. 3]{CS12}, we can observe that the load of small cells is much heavier than that of macrocells, hence resulting in small cells that are congested. Consequently, only the privileged users in the macrocell center achieve high data rates, while the other users are starved. To cope with this problem, in~\cite{QY12} a low-complexity distributed user association algorithm was proposed for maximizing the user data rate related utility, which was defined as a logarithmic function of the user data rate. Since the logarithm is a concave function and has diminishing returns, allocating more resources to an already well-served user has low priority, whereas providing more resources to users having low rates is desirable, thereby encouraging both load balancing and user fairness. In~\cite{QY12}, by relaxing the primal deterministic user association to a fractional association, the intractable primal combinatorial optimization problem was converted into a convex optimization scenario. By exploiting the convexity of the problem, a distributed user association algorithm was developed with the assistance of dual decomposition and the gradient descent method, which converged to the optimum solution under the guarantee of not exceeding a certain maximum discrepancy from optimality. We note that the convergence speed of the gradient descent method heavily depends on the particular choice of the step size. For the same problem formulation as in~\cite{QY12}, a coordinate descent method was proposed in~\cite{KS14} for providing a rigorous performance guarantee and faster convergence compared to the algorithm in~\cite{QY12}. {In~\cite{HZSM15}, the user association in femtocell networks was formulated as a combinatorial problem for minimization of the latency of service requested by the users, which was solved with the aid of approximation algorithms achieving a proven performance bound.}

Game theory is also widely applied in the context of user association for spectrum efficiency optimization. The downlink user association for HetNets was formulated as a bargaining problem in~\cite{DL14CCNC,DL14letter}, where the BSs acted as players competing for serving users. In~\cite{DL14CCNC}, a bargaining problem was formulated for the maximization of the data rate based utility, while guaranteeing a certain minimal rate for the users, and simultaneously maintaining fairness for all users as well as balancing the traffic load of the cells in different tiers. By extending the contribution of~\cite{DL14CCNC}, the QoS was maintained for multi-service traffic in~\cite{DL14letter}, where an opportunistic user association algorithm was developed for classifying human-to-human traffic as the primary service and machine-to-machine traffic as the secondary service. The proposed opportunistic user association aimed for providing fair resource allocation for the secondary service without jeopardizing the QoS of the primary service. Furthermore, the authors of~\cite{SO14,NN14,PF13} formulated the downlink user association in HetNets as a many-to-one matching game, where users and BSs evaluated each other based on well defined utilities. In~\cite{SO14}, users and BSs ranked one another based on specific utility functions that accounted for both the data rate and the fairness to cell edge users, which was captured in terms of carefully coordinated priorities. In contrast to~\cite{SO14}, which relied on differentiating user priorities, in~\cite{NN14} the delivery time, handover failure probability, and heterogeneous QoS requirements of users were taken into consideration when designing utility based user association. With the aid of a problem formulation similar to the one in~\cite{NN14}, the authors of~\cite{PF13} specifically focused their attention on multimedia data services and characterized the user's quality of experience in terms of mean opinion scores that accurately reflected the specific characteristics of the wireless application considered. Another interesting study was disseminated in~\cite{SW14}, where the uplink user association of HetNets was formulated as a college admission game combined with transfers, where a number of colleges, i.e., the BSs in macrocells and small cells, sought to recruit a number of students, i.e., users. The college admission game formulated carefully captures the users' need to optimize their packet success rates and delays, as well as the small cell's incentive to offload traffic from the macrocell and thereby to extend its coverage.

The densely deployed small cells further exacerbate the demand for interference management in HetNets. The joint optimization of user association and of other aspects of radio resource allocation understandably has prompted significant research efforts. In~\cite{FD11}, joint optimization of user association and channel allocation decisions between macrocells and small cells was investigated with the objective of maximizing the minimum data rate. Extending the advance proposed in~\cite{FD11}, in~\cite{GJ12} joint user association, transmission coordination, and channel allocation between macrocells and small cells was proposed for the sake of maximizing the data rate based utility. Joint user association and power control was investigated in the context of the downlink of HetNets in~\cite{KS14,MR10,SR15}, and for the uplink of HetNets in~\cite{VNH14,MH13}. The algorithms proposed in~\cite{KS14,MR10,SR15,VNH14} iteratively updated both the user association solution and the transmit power until convergence was attained. The authors of~\cite{MH13} formulated the sum throughput maximization problem as a non-cooperative game, with both users and BSs acting as players. {In \cite{MJF14}, a cooperative small cell network architecture was proposed, where both user association as well as spectrum allocation and interference coordination were implemented through the cooperation of neighbouring cells, so as to enhance the capacity of hotspots.}  We note that the aforementioned joint optimization of user association and channel allocation/power control turns out to be NP-hard, hence finding the optimal solution is not trivial. The solution may be approached for example by updating the user association and the power level sequentially in an iterative manner until convergence is reached as in~\cite{KS14,MR10,SR15,VNH14}. Alternatively, the user association may first be optimized with the aid of fixed channel allocation/transmission coordination and then followed by optimizing the channel allocation/transmission coordination accordingly and vice versa, as in~\cite{FD11,GJ12,MH13,MJF14}. As a result, we may conclude that careful user association optimization is crucial for the holistic optimization of HetNets, indisputably underlining the significance of a survey on user association.

The qualitative comparison of the above-mentioned user association algorithms for spectrum efficiency optimization in HetNets is detailed in Table~\ref{table:1}.

\subsection{User Association for Energy Efficiency Optimization}
The escalating data traffic volume and the dramatic expansion of the network infrastructure will inevitably trigger an increased energy consumption in wireless networks. This will directly increase the greenhouse gas emissions and mandate an ever increasing attention to the protection of the environment. Consequently, both industry and academia are engaged in working towards enhancing the network energy efficiency.

Maximizing the network energy efficiency may be supported by maximizing the amount of successfully sent data, while minimizing the total energy consumption. As far as the problem formulation is concerned, maximizing the network energy efficiency can be either expressed as minimizing the total energy consumption while satisfying the associated traffic demands or maximizing the ratio between the total data rate of all users and the total energy consumption of the network, which is defined as the overall energy efficiency (bits/Joule). Note that macrocells have a significantly higher transmit power than small cells, thus the access network energy consumption is typically higher when a user is associated with a macrocell. Hence, the network energy efficiency is crucially dependent on the user association decisions~\cite{MA14COM}.

Numerous valuable contributions have been published on energy efficient user association in HetNets~\cite{PH13,MA14ICC,HZ12,YRX14,LPQ13}. In~\cite{PH13}, a user association algorithm was developed for the uplink of HetNets in order to maximize the system energy efficiency subject to users' maximum transmit power and minimum rate constraints. In~\cite{MA14ICC}, user association for the downlink of HetNets was optimized by maximizing the ratio between the total data rate of all users and the total energy consumption. In contrast to the problem formulation in~\cite{MA14ICC}, in~\cite{HZ12} the authors investigated energy efficient user association by minimizing the total power consumption, while satisfying the users' traffic demand. The authors of~\cite{YRX14} considered the association problem for users involving video applications, where a video content aware energy efficient user association algorithm was proposed for the downlink of HetNets, with the goal of maximizing the ratio of the peak-signal-to-noise-ratio and the system energy consumption. Thereby, both nonlinear fractional programming and dual decomposition techniques were adopted for solving the problem. In~\cite{LPQ13}, a Benders' decomposition~\cite{bender} based algorithm was developed for joint user association and power control with the goal of maximizing the downlink throughput. This was achieved by associating every user with the specific BS, which resulted in the minimization of the total transmit power consumption.

Statistical studies of mobile communication systems have shown that 57\% of the total energy consumption of wireless networks can be attributed to the radio access nodes~\cite{YC11}. Furthermore, about 60\% of the power dissipated at each BS is consumed by the signal processing circuits and air conditioning~\cite{AG11}. As a result, shutting down BSs which support no active users is believed to be an efficient way of reducing the network power consumption~\cite{DF13,RJB14}.

In~\cite{LS13}, joint optimization of the long-term BS sleep-mode operation, user association, and subcarrier allocation was considered for maximizing the energy efficiency or minimizing the total power consumption under the constraints of maintaining an average sum rate target and rate fairness. The performance of these two formulations (namely, energy efficiency maximization and total power minimization) was investigated using simulations. In~\cite{CRE15}, an energy efficient algorithm was introduced for minimizing the energy consumption by beneficially adjusting both the user association and the BS sleep-mode operations, where the dependence of the energy consumption both on the spatio-temporal variations of traffic demands and on the internal hardware components of BSs were considered. Additionally, in \cite{YSS13} the coverage probability and the energy efficiency of K-tier heterogeneous wireless networks were derived under different sleep-mode operations using a stochastic geometry based model. The authors formulated both power consumption minimization as well as energy efficiency maximization problems and determined the optimal operating regimes of the macrocell.

The qualitative comparison of the above-mentioned user association algorithms for energy efficiency optimization in HetNets is detailed in Table~\ref{table:1}.

\subsection{User Association Accommodating Other Emerging Issues in HetNets}
Apart from the transmit power disparity between small cells and macrocells in HetNets, the inherent nature of HetNets manifests itself in terms of the uplink-downlink asymmetry, the backhaul bottleneck, diverse footprints and so on. This imposes substantial challenges for the user association design. However, these issues have only briefly been alluded to in the majority of the existing research. In the following, we highlight three crucial issues, as summarized in Table~\ref{table:2}.

\begin{table*}[t!]
\newcommand{\tabincell}[2]{\begin{tabular}{@{}#1@{}}#2\end{tabular}}
  \centering
   \caption{Summary of Emerging Issues in User Association for HetNets}
  \begin{tabular}{|l|r|r|r|}\hline
\textbf{Issues} &\textbf{Example} & \textbf{Key point}& \textbf{Ref.} \\ \hline
\tabincell{l}{Uplink-downlink\\ asymmetry} & \tabincell{r}{HetNets introduce a major asymmetry between\\ the uplink and the downlink. The optimal \\user association for downlink or uplink will \\be less effective for the opposite direction.} & \tabincell{r}{Optimize downlink and uplink performance\\jointly in the user association design.}&~\cite{XC12,TZYH15,DL14VTC,SL15}\\ \hline
\tabincell{l}{Backhaul bottleneck} & \tabincell{r}{Densely deployed small BSs may introduce \\overwhelming traffic augments for the backhaul\\link and current small cell backhaul solutions\\ cannot provide sufficiently large data rate.} & \tabincell{r}{Design backhaul-aware user association for\\ HetNets.} & ~\cite{BH15,NW15,ZC14,DD13,PFBM14,FR13,WCL14,MA14COM}\\ \hline
\tabincell{l}{Mobility support} & \tabincell{r}{User association without considering\\user mobility may result in frequent \\handovers among the cells in HetNets.} & \tabincell{r}{Account for the user mobility when making the user\\ association decision in HetNets to enhance the long-term \\system-level performance and avoid excessive handovers.}& ~\cite{SS15,CP13,MS14,DCOT12,DHAHRI2014,WRP15,MK15}\\ \hline
\end{tabular}
\label{table:2}
\end{table*}

\subsubsection{Uplink-Downlink Asymmetry}
Most of the research on user association in HetNets investigated the problem from either a downlink or an uplink perspective. However, HetNets typically introduce an asymmetry between uplink and downlink in terms of the channel quality, the amount of traffic, coverage, and the hardware limitations. Amongst them, the uplink and downlink coverage asymmetry is the more severe in HetNets. In the downlink, due to the large power disparities between the different BS types in a HetNet, macrocells have much larger coverage areas than small cells. By contrast, the users' devices may transmit at the same power level in the uplink,  regardless of the BS type. Although some promising results related to decoupling of the uplink and downlink user association have been reported in~\cite{KSHE14,Smiljkovikj_K2015,SSXZ14}, this decoupling inevitably requires a tight synchronization as well as a high-speed and low-delay data connectivity between BSs. Ever since the inception of mobile telephony, users have been constrained to associate with the same BS in both the downlink and uplink directions, since this coupling makes it easier to design and operate the logical, transport, and physical channels~\cite{Federico_Boccardi2015}. Hence, a coupling of uplink and downlink may also be expected for 5G networks. Due to the uplink-downlink coverage asymmetry of HetNets, a user association that is optimal for either the downlink or the uplink may become less effective for the opposite direction. Specifically, the max downlink RSS based user association rule may associate a user with the far-away marcocell, rather than with the nearby small cell. As a result, the user has to transmit at a potentially excessive power for guaranteeing the target received signal strength in the uplink, thereby inflicting a high uplink interference on the small cell users, hence degrading both the spectrum and energy efficiencies as well as shortening the battery recharge period. {We note that the uplink-downlink asymmetry exists, regardless of whether time-division duplex (TDD) or frequency-division duplex (FDD) is adopted for separating the uplink and downlink transmissions in HetNets.}

Consequently, in HetNets using sophisticated joint uplink and downlink user association optimization is imperative. Hence, the authors of~\cite{XC12} proposed a user association algorithm, which jointly maximized the number of users admitted and minimized the weighted total uplink power consumption. However, the performance of the algorithm highly depended on the specific weight, which was heuristically obtained in~\cite{XC12}. The objective function used in~\cite{XC12} was further improved in~\cite{TZYH15} so as to maximize the network utility, which was based on the ratio between the downlink data rate and the uplink power consumption. In~\cite{DL14VTC}, the user association optimization problem was formulated as a bargaining problem configured for maximizing the sum of uplink and downlink energy efficiency related utilities. In~\cite{SL15}, a user association and beamforming algorithm was developed for minimizing the total uplink and downlink energy consumption, under specific QoS constraints for the users.

\subsubsection{Backhaul Bottleneck}
HetNets are expected to constitute a cellular paradigm shift, which raises new research challenges. Among these challenges, the importance of the backhaul bottleneck has not been fully recognized in the context of the 4G LTE network~\cite{AJG13}. Specifically, most of the research assumed a perfect backhaul between the BS and the network controller, and focused on the achievable performance gains of the wireless front-end without taking into account the specific details of the backhaul implementation and any possible backhaul bottleneck. This assumption is generally correct for well-planned classical macrocells. However, in HetNets the potentially densely deployed small BSs may impose an overwhelming backhaul traffic. On the other hand, the current small cell backhaul solutions, such as xDSL and non-line-of-sight (NLOS) microwave, are far from an ideal backhaul solution owing to their limited data rate~\cite{NTNS13}. As already observed in~\cite{BN14}, the full benefits of dense HetNets can be realized only if they are supported by the careful consideration of the backhaul. {Hence, the backhaul capacity constraint is of considerable importance in HetNets.}

{Therefore, for HetNets, backhaul-aware user association mechanisms, which fully take the backhaul capacity constraint into account, are needed.} A distributed user association algorithm was developed for maximizing the network-wide spectrum efficiency in~\cite{BH15} involving relaxed combinatorial optimization. Similarly, a sum of the user rate based utility was investigated in~\cite{NW15} under backhaul constraints. In~\cite{ZC14}, a waterfilling-like user association algorithm was devised for maximizing the weighted sum rate of all users in conjunction with carrier aggregation (CA), while enforcing a particular backhaul constraint for small cell BSs. Furthermore, the authors of~\cite{DD13} conceived a heuristic user association algorithm for maximizing the overall network capacity under both backhaul capacity and cell load constraints. In~\cite{PFBM14}, cache-aware user association was designed using the power of game theory in backhaul-constrained HetNets, which was modeled as a one-to-many matching game. Specifically, the users' and BSs' association were characterized based on the capacity and by giving cognizance to the utility that accounted for both the BSs' data storage capabilities and the users' mobility patterns. The authors of \cite{FR13} presented an intriguing model, where third parties provided the BSs with backhaul connections and leased out the excessive capacity of their networks to cellular providers, when available, presumably at a significantly lower cost than that of QoS guaranteed connections. The authors provided a general user association optimization algorithm that enabled the cellular provider to dynamically determine which specific users should be assigned to third-party femtocells based on the prevalent traffic demands, interference levels as well as channel conditions and third-party access pricing. With user fairness in mind, the authors of~\cite{WCL14} considered the joint resource allocation across wireless links and the flow control within the backhaul network for maximizing the minimum rate among all users. Turning our attention to energy efficiency, Mesodiakaki \emph{et al.}~\cite{MA14COM} studied energy efficient user association issues of HetNets by taking both the access network's and the backhaul's energy consumption into account.

\subsubsection{Mobility Support}
The increased cell densification encountered in HetNets continuously poses challenges for mobility support. The reduced transmit powers of small cells lead to reduced footprints. As a result, for a user having moderate or high mobility, a user association algorithm that does not consider the mobility issues may result in more frequent handovers among the cells in HetNets compared to conventional homogeneous cellular networks. However, it is well understood that handovers trigger a whole host of complex procedures, impose costly overheads as well as undesirable handover delays and possibly dropped calls. Moreover, as shown in a 3GPP technical report~\cite{3gppMobility}, the handover performance experienced in HetNets is typically not as good as that in systems with pure macrocell deployment. Hence, it is imperative to account for user mobility, when making user association decisions in HetNets in order to enhance the long-term system-level performance and to avoid excessive handovers.

Taking advantage of stochastic geometry, the authors of~\cite{SS15} first derived the downlink coverage probability considering the users' speed, under the biased user association rule. Then, the optimal bias maximizing the coverage was obtained, where both the optimal bias and the coverage probability were related to the users' speed. Not surprisingly, the results in~\cite{SS15} revealed that a speed-dependent bias factor was capable of effectively improving both the coverage probability and the overall network performance. In~\cite{CP13}, the authors modeled the user mobility by a Markov modulated Poisson process~\cite{MMPP} and jointly considered it with the user association problem with the goal of optimizing the system performance in terms of the average traffic delay and the blocking probability. Moharir \emph{et al.}~\cite{MS14} studied user association in conjunction with the effect of mobility in two-tier downlink HetNets and showed that traditional algorithms that only forwarded each packet at most once either to a BS or to a mobile user had a poor delay performance. This unexpected trend prevailed, because the rapidly fluctuating association dynamics between BSs and users necessitated a multi-point relaying strategy, where multiple BSs stored redundant copies of the data and coordinated for reliably delivering the data to mobile users. {The authors of~\cite{DCOT12,DHAHRI2014} provided an interesting framework, where Q-learning was adopted for finding the best user association in a non-stationary femtocell network by exploring the past cellular behavior and predicting the potential future states, so as to minimize the frequency of handovers.}

{Recently, dual connectivity has been standardized in 3GPP release 12~\cite{3gppDC13,3gppDC14,DCwhite} as a remedy for  mobility support in HetNets. Dual connectivity can significantly  improve the mobility resilience and increase the attainable user throughput due to its potential of extending the CA and coordinated multi-point (CoMP) to multiple BSs, which allows users to be simultaneously associated with both macro BSs and small cell BSs. More specifically, dual connectivity enables a user to maintain the connection to the macro BS and receive signalling messages as long as it is in its coverage area. This user does not have to initiate handover procedures unless moving to the coverage of another macro BS, thereby indisputably handling the handover more efficiently. User association, which determines the specific BSs the user should associate with, pre-determines the relative performance gain achievable with dual connectivity. In~\cite{WRP15}, the impact of different user association criteria on the attainable  performance of dual connectivity was evaluated via simulations. In~\cite{MK15}, the user association problem for  maximization of the sum rate of all users was formulated, and a low-complexity sub-optimal user association algorithm was proposed  for solving the problem. It is noted that despite the potential benefits of dual connectivity in enhancing the resilience to user mobility and increasing the user throughput, dual connectivity also imposes several technical challenges in terms of buffer status report calculation and reporting, transmission power management, etc.~\cite{JSC14}. As such, more efforts have to be devoted to tackling these issues before fully deploying dual connectivity for enhanced mobility support in HetNets.}

\subsection {Summary and Discussions}
Supporting QoS, spectrum efficiency, energy efficiency and fairness in 5G networks is an essential requirement for real-time applications. How to address these performance metrics at the user association stage is becoming increasingly important~\cite{HRQ14}. From Table~\ref{table:1}, we observe that most existing research contributions do not take all of these metrics into account.

A theoretical analysis of the tradeoffs between the energy efficiency and spectrum efficiency under the additional consideration of the user QoS and fairness was carried out for downlink orthogonal frequency-division multiple access (OFDMA) scenarios in~\cite{CX11} and for homogeneous cellular networks in~\cite{RJB13}. A similar tradeoff is expected for HetNets. However, how to characterize this tradeoff with closed-form expressions remains an open challenge, since HetNets are much more complex than conventional homogeneous networks. As such, {more research is required for theoretically analyzing the tradeoff amongst the attainable spectrum efficiency, energy efficiency, QoS, and fairness in HetNets, which can provide deep engineering insights regarding the interplay of these performance metrics. This could provide guidelines for conceiving user association strategies that simultaneously cater to all of these performance metrics.}

Apart from the above-mentioned interplay of the performance metrics, the inherent nature of HetNets has imposed a plethora of challenging issues, such as the uplink and downlink asymmetry, the backhaul bottleneck, and the need for efficient mobility support. All these issues have been set aside for future research by the majority of the existing works. Although there is some initial research addressing each one of the above issues in the context of user association algorithm design as shown in Table~\ref{table:2}, these topics are still in their infancy and more theoretical analysis as well as practical independent verification is required.

{Finally, most of the existing contributions on user association conceived for HetNets focus on either optimisation or theoretical performance analysis. The application of theoretical results to realistic models and practical systems is still an open area. To make further progress, conceiving a holistic architecture, which employs the aforementioned advanced technologies to provide improved QoS, spectrum efficiency, energy efficiency, and fairness, as well as accounting for the issues imposed by the inherent nature of HetNets becomes highly desirable for 5G evolution.}

\section{User Association in Massive MIMO Networks}
Massive MIMO~\cite{Marzetta2010} constitutes a fundamental technology for achieving the ambitious goals of 5G systems and hence it has attracted substantial interests from both academia and industry. This new design paradigm may be viewed as a large-scale multi-user MIMO technology, where each BS is equipped with a large antenna array and communicates with multiple terminals over the same time and frequency band by distinguishing the users with the aid of their unique user-specific channel impulse response (CIR)~\cite{Emil_BJ_10myths_2015}. Compared to current small scale MIMO networks, massive MIMO systems achieve high power and spectrum efficiencies, despite their low-complexity transceiver designs. Random impairments, such as small-scale fading and noise, are averaged out, when a sufficiently high number of antennas are deployed at the BS~\cite{ngo2013energy}. Moreover, the effects of interference, channel estimation errors, and hardware impairments~\cite{E_Bjornson2014} vanish, when the number of antennas becomes sufficiently high, leaving only the notorious pilot contamination problem as the performance-limiting factor~\cite{Hoydis2013}. The implementation of massive MIMO is also beneficial in other networks, such as cognitive radio networks. It was shown in \cite{Lifeng_massiveMIMO} that when the number of primary users was proportional to the number of antennas at the primary BS, the number of antennas at the secondary BS should be larger than the logarithm of the number of primary users, in order to mitigate the effects of interference.

The distinct characteristics of massive MIMO inevitably necessitate the redesign of user association algorithms. On the one hand, BSs equipped with large antenna arrays provide high multiplexing gains and array gains. On the other hand, the power consumption increases due to the more complex digital signal processing. In the following, we investigate the effects of massive MIMO on user association in terms of the received power, throughput, and energy efficiency. Table~\ref{Table_MassiveMIMO} illustrates the qualitative comparison of existing user association algorithms conceived for massive MIMO networks.

\begin{table*}[tb]
\newcommand{\tabincell}[2]{\begin{tabular}{@{}#1@{}}#2\end{tabular}}
\renewcommand{\multirowsetup}{\centering}
\scriptsize
\centering
  \caption{Qualitative Comparison of User Association Algorithms for Massive MIMO Networks.}
\begin{tabular}{|l|r|r|r|r|r|r|r|r|r|r|}
  \hline
  \textbf{Ref.}&\textbf{Algorithm}&\textbf{Topology }&\textbf{Model }& \textbf{Direction}&\textbf{Control }& \textbf{\tabincell{r}{Spectrum\\efficiency}}& \textbf{\tabincell{r}{Energy\\efficiency}}& \textbf{\tabincell{r}{QoS\\provision}}& \textbf{Fairness}&\textbf{\tabincell{r}{Coverage\\probability}}\\
  \hline
   \cite{AnqiHe_2015} & max-RSS UA & \multirow{2}{1.3cm}{Random spatial} & \multirow{2}{1.3cm}{Stochastic geometry} & DL & Distributed& Low& -& Moderate& -& Low\\
   \cite{Anqi_COMML_2015} & biased UA & &  & DL & Distributed & High & High & -& -& Moderate\\
  \hline
  \cite{DBLP2014} & UA & \multirow{2}{0.8cm}{Grid} &  \multirow{2}{0.8cm}{Game theory} & DL & Distributed & High& - & Moderate & High& -\\
  \cite{Y_Xu_2015_massiveMIMO} & UA & & & DL & Centralized/Distributed & High & - & Moderate & High &-\\
  \hline
  \cite{NW15} & UA &  \multirow{2}{0.8cm}{Grid}& \multirow{2}{1.3cm}{Combinatorial optimization}& DL & Centralized& High& -&High & High &- \\
  \cite{Dantong_massivMIMO_2015} & UA& & & DL & Distributed & -& High & High & High &-\\
  \hline
\end{tabular}
\label{Table_MassiveMIMO}
\end{table*}

\subsection{Received Power Based User Association}
In the massive MIMO downlink, the $N$-antenna BS typically uses linear transmit pre-coding (TPC) schemes to transmit data signals to $S$ users relying on the knowledge of the downlink CIR, which facilitates the use of a low-complexity single-user receiver. There are two commonly used TPC schemes, i.e., maximum ratio transmission (MRT) and zero-forcing (ZF) TPC~\cite{EG_Larsson_COMMAG_2014}. For MRT, the  power received  at the user is proportional to $N$. For ZF TPC, the intra-cell interference is cancelled at the cost of reducing the degrees of freedom and hence the diversity order to $\left(N-S+1\right)$~\cite{Hosseini2014_Massive}. For instance, by using ZF TPC with equal power allocation, the long-term average  power received  at the intended user can be expressed as
\begin{align}\label{AG_MassiveMIMO}
P_r=\left(N-S+1\right) \cdot \frac{P_t}{S} \cdot L,
\end{align}
where  ${P_t}$ is the BS's transmit power and $L$ is the path-loss. Eq. \eqref{AG_MassiveMIMO} reveals that for user association based on the maximum received power, the cell coverage is expanded due to the large antenna array gain.

\begin{figure}[t!]
        \center
        \includegraphics[width=3.8 in]{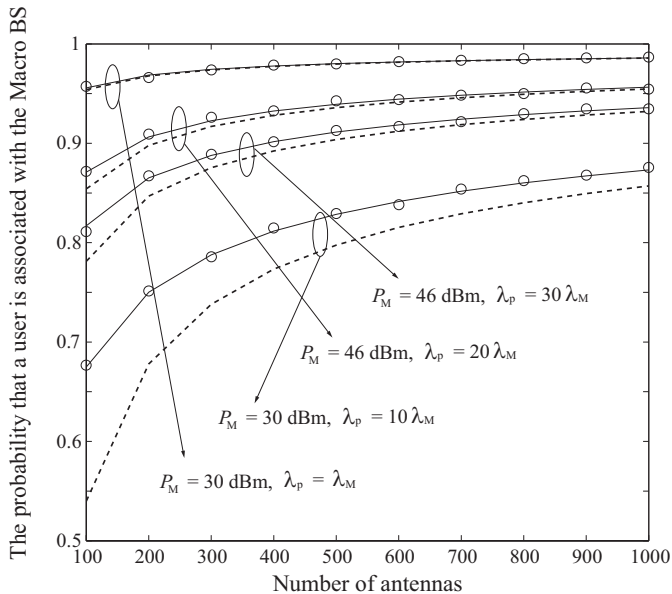}
        \caption{The probability that a user is associated with the macro BS. The locations of the macro BSs and the pico BSs follow independent homogeneous Poisson point processes with densities $\lambda_\mathrm{M}$ and $\lambda_\mathrm{p}$, respectively. The carrier frequency is $1$ GHz, $S=5$, $P_\mathrm{M}$ is the macro BS transmit power, and $P_\mathrm{p} = 30$ dBm is the pico BS transmit power. The solid lines are validated by Monte Carlo simulations marked with `o' and the dashed lines represent the asymptotic results as the number of antennas goes towards infinity.}
        \label{massiveMIMO_UserA}
\end{figure}

In HetNets, massive MIMO may be adopted in macrocells, since the physically large macro BSs  can be readily equipped with large antenna arrays~\cite{Jungnickel_IEEE_Commag}. In~\cite{AnqiHe_2015}, a stochastic geometry based approach is invoked for analyzing the impact of massive MIMO on the max-RSS user association. Considering a two-tier HetNet consisting of macrocells and picocells, Fig. \ref{massiveMIMO_UserA} shows the probability that a user is associated with the macro BS relying on the max-RSS user association. We observe that even when the macro BS reduces its transmit power to the
same level as the pico BS  (i.e., $P_{\rm{M}}=30 $ dBm), a user is still much more likely to be associated with the macro BS than with the pico BS, due to the large array gain brought by the massive MIMO macro BS. We also observe that increasing the number of transmit antennas at the macro BS improves the probability that a user is associated with the macro BS, which indicates that macro BSs having large antenna arrays are capable of carrying higher traffic loads, hence reducing the number of small cells required.

\subsection{User Association for Spectrum Efficiency Optimization}
Massive MIMO is capable of achieving a high spectrum efficiency by simultaneously transmitting/receiving multiple data streams  in the same band. The low-complexity max-RSS user association may  be incapable of balancing the load in multi-tier HetNets with the aid of massive MIMO~\cite{D_Bethanabhotla_2014_1}, as indicated in Fig. \ref{massiveMIMO_UserA}.
In \cite{D_Bethanabhotla_2014_1},  load balancing was investigated in multi-tier networks where the BSs of different tiers were equipped with different numbers of antennas and used linear ZF beamforming (LZFBF) for communicating with different numbers of users.  Bethanabhotla \emph{et al.}~\cite{D_Bethanabhotla_2014_1} focused their attention on a pair of user-centric association algorithms, namely on the max-rate association and on the load based association. Under max-rate association, each user decided to associate with the specific BS which provided the maximum peak rate. In contrast, for load based association, each user aimed for selfishly maximizing its own throughput by also considering the traffic load of the BS. The performance of these user centric association schemes was also examined in \cite{D_Bethanabhotla_2014_1}.  In~\cite{DBLP2014}, a user-centric distributed probabilistic scheme was proposed for massive MIMO HetNets, which showed that the proposed scheme converged to a pure-strategy based Nash equilibrium with a probability of one for all the practically relevant cases of proportional fair and max-min fair utility functions. In \cite{Y_Xu_2015_massiveMIMO}, user association was investigated in two-tier HetNets with a massive MIMO aided macrocell and multiple conventional picocells.  Both the centralized and distributed perspectives were considered. The goal of Gotsis \emph{et al.}~\cite{A_G_Gotsis_2015ICC} was to identify the optimal user-to-access point association decision for maximizing the worst rate in the entire set of all users in massive MIMO empowered ultra-dense wireless networks. This contribution showed that at the network-level, optimal user association designed for densely and randomly deployed massive MIMO networks had to account for both the channel and traffic load conditions. In \cite{NW15}, joint downlink user association and wireless backhaul bandwidth allocation was considered for a two-tier HetNet, where small cells relied on in-band wireless links connecting them to the massive MIMO macro BS for backhauling. It was shown that under the specific wireless backhaul constraint considered, the joint scheduling problem for maximizing the sum of the logarithm of the rates for users constituted a nonlinear mixed-integer programming problem. The authors of \cite{NW15} also considered the global and local backhaul bandwidth allocation.

\subsection{User Association for Energy Efficiency Optimization}
The energy efficiency of massive MIMO systems has been  studied in~\cite{ngo2013energy,Hong_yang_2013,BE2013_ICT,BE2014_WCNC,E_Bjornson2014,Emil_JSAC_2015}. In \cite{ngo2013energy}, the energy efficiency and spectrum efficiency tradeoff was analyzed. However, Ngo \emph{et al.} \cite{ngo2013energy} only took into account the transmit power consumption, but not the circuit-power dissipation. In practice, the internal non-RF power consumption scales with the number of antennas~\cite{Hong_yang_2013}. Compared to a typical LTE BS, it is implied in \cite{Hong_yang_2013} that BSs with large scale antennas achieve much higher energy efficiencies.  In \cite{BE2014_WCNC}, a more specific power consumption model was provided to show how the power scales with the number of antennas and the number of users. In this power consumption model, both the RF power consumption and the circuit power consumption including the digital signal processing and the analog filters used for RF and baseband processing were considered. An important insight obtained from \cite{BE2014_WCNC} is that although using hundreds of antennas expends more circuit-power, the per-user energy efficiency still improves by serving an increased number of  users with interference-suppressing regularized ZF TPC.

Energy-efficient user association in massive MIMO aided HetNets is still in its infancy. In \cite{Anqi_COMML_2015}, the impact of flexible user association on the energy efficiency in K-tier massive MIMO enabled HetNets was evaluated. In \cite{BE2013_ICT}, soft-cell coordination was investigated where each user could be served by a dedicated user-specific beam generated via non-coherent beamforming, and the total power consumption was minimized under a specific QoS constraint. In \cite{Emil_JSAC_2015}, the uplink energy efficiency of a massive MIMO assisted cellular network was analyzed, where stochastic geometry was applied for modeling the network and the user association was based on the minimum path-loss criterion. Distributed energy efficient and fair user association in massive MIMO assisted HetNets was first proposed in our recent work~\cite{Dantong_massivMIMO_2015}, where user association objective was to maximize the geometric mean of the energy efficiency, while considering QoS provision for users.
  \begin{figure} [tbp]
\vspace{-0.cm}
    \centerline{\includegraphics[width=3.8 in]{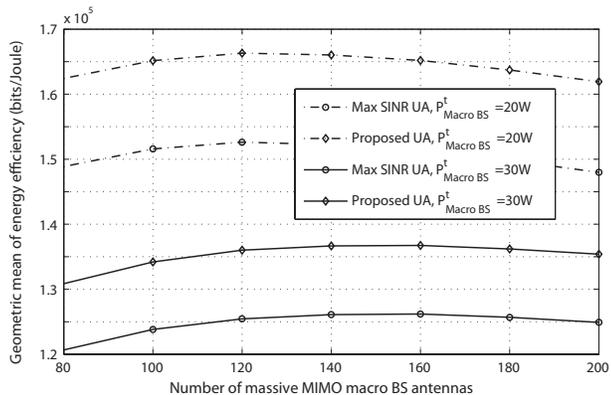}}
    \caption{Energy efficiency versus the number of antennas for different downlink transmit powers of the macro BS {\small{$P_{\mathrm{Macro\;BS}}^\mathrm{t}$}}.}
    \label{fig:2}
\end{figure}
Fig.~\ref{fig:2} illustrates the geometric mean of the energy efficiency of different numbers of antennas and downlink transmit powers at the macro BS. We observe that regardless of both the number of antennas and the transmit power of the macro BS, our proposed algorithm achieves a better energy efficiency than the max SINR based user association algorithm. For a given macro BS's transmit power, the energy efficiency initially slightly increases and then gracefully decreases with increasing number of antennas. This is attributed to the fact that when the number of antennas exceeds a critical value (approximately 120 when the transmit power of macro BS is 20 W, {\small{$P_{\mathrm{Macro\;BS}}^\mathrm{t}=20$ W}}), adding more antennas improves spectrum efficiency, but as usual, at the cost of degrading energy efficiency due to the increased power consumption. Fig.~\ref{fig:2} also illustrates that given the same number of antennas, a lower macro BS transmit power facilitates a higher energy efficiency. This trend demonstrates the superiority of massive MIMO in fulfilling the QoS requirement at a reduced transmit power.
\begin{figure} [tbp]
    \centerline{\includegraphics[width=3.8 in]{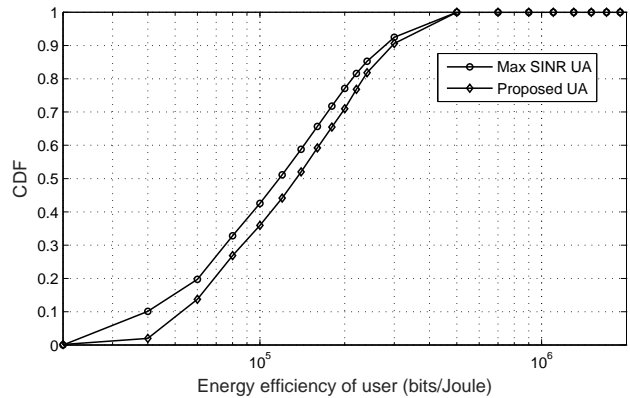}}
    \caption{Cumulative distribution function of user energy efficiency.}
    \label{fig:4}
\end{figure}
To provide further insights, Fig.~\ref{fig:4} shows the cumulative distribution function (CDF) of the energy efficiency experienced by users expressed in bits/Joule for different user association algorithms. We set the macro BS's downlink transmit power to 30W and the number of macro BS antennas to $N=100$. We observe that the proposed algorithm improves the CDF in the low energy efficiency domain. The CDF of max SINR based user association approaches that of our proposed algorithm at an energy efficiency of $4 \times 10^5$ bits/Joule. This can be explained by the fact that, as the objective of the proposed algorithm, maximizing the geometric mean of energy efficiency leads to  proportional fairness, which provides a more uniform energy efficiency by reassigning resources from the users. As such, our proposed algorithm~\cite{Dantong_massivMIMO_2015} improves the user fairness in terms of the energy efficiency compared to the max SINR  based user association algorithm.

\subsection{Summary and Discussions}
The application of massive MIMO has a substantial effect on user association. The existing contributions summarized in Table~\ref{Table_MassiveMIMO} have shown the importance of user association in massive MIMO aided networks. User association schemes designed for the already operational cellular systems may not be capable of fully exploiting the specific benefits provided by massive MIMO BSs. For the emerging 5G HetNets employing massive MIMO, the design of new user association schemes  is required, and there are at least two aspects that should be taken into account: 1)  The max-RSS based user association may force the massive MIMO BS to carry most of data traffic in HetNets, resulting in significant load imbalance between the macrocells and picocells. Therefore, throughput load balancing is important in massive MIMO assisted HetNets; and 2) Although massive MIMO uses large numbers of antennas and requires more power for complex signal processing, it can still remain energy efficient by serving more users since the power consumption per user is reduced and an increased spectrum efficiency is achieved. As such, energy efficient user association in massive MIMO HetNets should carefully balance the interplay between the number of antennas at the BS and the number of users served by the massive MIMO BS.

From the discussions above, we conclude that user association in massive MIMO HetNets is a promising research topic, and hence more research efforts are needed for facilitating its enhancement in practical 5G networks.

\section{User Association in mmWave Networks}
Due to the rapid proliferation of bandwidth-hungry mobile applications, such as video streaming with up to high definition television (HDTV) resolution, more spectral resources are required for 5G mobile communications~\cite{maged_editoral1}. However, the existing cellular band is already heavily used and even using CA~\cite{TR2013} relying on several parallel carries fails to meet the high spectrum demand of 5G networks. New spectrum has to be harnessed at higher carrier frequencies. Hence, mmWave communications with a large bandwidth have emerged as a potentially promising 5G technology~\cite{ZP2011,TSR2013}.

\subsection{MmWave Channel Characteristics}
The channel quality between the user and the BS plays a key role in user association. Hence, we first have to understand the mmWave channel. The mmWave channel characteristics can be highlighted as follows:
\begin{itemize}
  \item \textbf{Increased path-losses}. According to the Friis transmission formula~\cite{Friis_1946} of free-space propagation, the path-loss increases with the square of the carrier frequency, which indicates that mmWave transmissions suffer from high power losses.

  \item \textbf{Different propagation laws}. NLOS signals suffer from a higher attenuation than LOS signals~\cite{TSR2013}. This important feature of the propagation environment has to be incorporated into the design and analysis of mmWave networks~\cite{T_Bai2014}.

  \item \textbf{Sensitivity to blockages}. MmWave signals are more sensitive to blockage effects than signals in lower frequency bands, and indoor users are unlikely to be adequately covered by outdoor mmWave BSs~\cite{T_Bai2014}.
\end{itemize}
These channel characteristics have a significant effect on cell coverage, which indicates that the attainable throughput of mmWave networks is highly dependent on the user association~\cite{JZ2014_GC_Workshop}. In~\cite{T_Bai2014,S_Singh2014}, the users are assumed to be associated with the specific BS offering the minimum path-loss, where stochastic geometry based mmWave modeling incorporating blockage effects was considered. It was demonstrated in~\cite{T_Bai2014,S_Singh2014} that mmWave networks tended to be noise-limited, because the high path-loss attenuated the interference, which was likely to be further attenuated by directional beamforming. {Hence, user association metrics designed for interference-limited homogenous systems are not well suited to mmWave systems~\cite{HG2015}. In contrast, user association should be designed to meet the dominant requirements of throughput and energy efficiency without considering interference coordination.} {Additionally, user orientation has a substantial impact on the performance of  mmWave links due to the fact that  directional transmission is required for combatting the high path-loss. As such, users may not be associated with the geographically closest BS, since a better directional link may exist for a farther away BS.}

\subsection{MmWave User Association}

Current standards for mmWave communications, such as the IEEE 802.11ad and IEEE 802.15.3c, adopt RSS-based user association, which may lead to an inefficient use of resources~\cite{Y_Niu2015}. Moreover, RSS-based user association may result in overly frequent handovers between the adjacent BSs and  may increase the overhead/delay of re-association~\cite{HG2015}. In~\cite{AG_2013}, the optimal assignment of the BSs to the available access points (APs) in 60 GHz mmWave wireless access networks was investigated, and a BS association method was proposed for maximizing the total requested throughput. Since the problem considered in~\cite{AG_2013} was a classical multi-assignment optimization problem where an AP was assigned to more than one BS, an auction-based solution was proposed. Xu \emph{et al.} \cite{YX2013} extended the line of work in~\cite{AG_2013} by deploying relays in the network, which helped the clients associated with the AP. More explicitly, a combination of distributed auction algorithms was used for jointly optimizing the client association and relay selection processes. In~\cite{AG2014_1}, the optimization of user association was carried out by giving special cognizance to both load balancing and fairness in mmWave wireless networks. The study in~\cite{AG2014_1} aimed for balancing the AP utilization in the network, in an effort to improve the throughput and fairness in resource sharing.

MmWave cells may be regarded as another tier in future 5G HetNets~\cite{ZP2011,KeiSakaguchi_2015,Y_Niu2015}. Unlike conventional HetNets,  where all the tiers use the same frequency band, which  necessitates interference management\cite{3GPP_TS36_300}, the mmWave tier has no effect on the other tiers, since it operates  at higher frequencies. Therefore, the deployment of mmWave cells not only offloads the data traffic of existing HetNets, but also reduces the interference by avoiding the deployment of cellular BSs in the cellular band. In \cite{KeiSakaguchi_2015}, multi-band HetNets conceived for 5G were considered, where the different tiers operated at different frequencies.  The 60 GHz band has a factor 100  more bandwidth than the current cellular bands.  To allow small cell BSs to accommodate more data traffic  and to maximize the system's data rate, a novel user association method using combinatorial optimization was introduced in \cite{KeiSakaguchi_2015}, where the supported achievable rate and the number of users in each cell  were considered. In \cite{Bingyu_Xu_2016}, user association was considered in a hybrid HetNet, where macro cells adopt massive MIMO, and small cells adopt mmWave transmissions. The work of \cite{Bingyu_Xu_2016} showed that the proposed algorithm can well coordinate the capabilities of massive MIMO and mmWave in the 5G networks.

Table~\ref{Table_mmWave} qualitatively compares the existing user association algorithms proposed for mmWave networks.

\begin{table*}[tb]
\newcommand{\tabincell}[2]{\begin{tabular}{@{}#1@{}}#2\end{tabular}}
\renewcommand{\multirowsetup}{\centering}
\scriptsize
\centering
  \caption{Qualitative Comparison of User Association Algorithms for mmWave Networks.}
\begin{tabular}{|l|r|r|r|r|r|r|r|r|r|r|}
  \hline
  \textbf{Ref.}&\textbf{Algorithm}&\textbf{Topology }&\textbf{Model }& \textbf{Direction}&\textbf{Control }& \textbf{\tabincell{r}{Spectrum\\efficiency}}& \textbf{\tabincell{r}{Energy\\efficiency}}& \textbf{\tabincell{r}{QoS\\provision}}& \textbf{Fairness}&\textbf{\tabincell{r}{Coverage\\probability}}\\
  \hline
   \cite{T_Bai2014,S_Singh2014} & max-RSS UA & Random spatial & Stochastic geometry & DL & Distributed& Low& -& Moderate& -& Low\\
  \hline
  \cite{AG_2013} & UA & Grid &  Game theory &  DL & Distributed & High& - &- & -& -\\
   \hline
  \cite{YX2013} & UA &\multirow{4}{0.8cm}{Grid} & \multirow{4}{1.3cm}{Combinatorial optimization}&  DL & Distributed & High & - & - & - &-\\
    \cite{AG2014_1} & UA &  & & DL& Distributed& High& -&- & High &- \\
  \cite{KeiSakaguchi_2015} & UA & & & DL & Centralized & High & -& High & High &-\\
  \cite{Bingyu_Xu_2016} & UA & & & DL & Distributed & High& High& -&-&-\\
  \hline
\end{tabular}
\label{Table_mmWave}
\end{table*}

\subsection{Summary and Discussions}
Although the aforementioned literature has shown the significant impact of user association on mmWave networks, user association in mmWave networks is far from being well understood and hence faces prominent challenges. RSS-based user association may not be feasible in future networks employing multiple frequency bands. The solution provided in \cite{AG_2013,YX2013,AG2014_1} ignored some important mmWave channel characteristics, such as the NLOS/LOS propagation laws and blockage. The velocity of mobility also has a substantial effect on mmWave user association/re-association, which suggests that  mobility management has to be adopted in mmWave networks~\cite{HG2015}. { Considering the fact that mmWave links are sensitive to blockage and mobility, the channel conditions may vary significantly over time and it may be necessary to request re-association after each channel coherence time.} {In addition, since mmWave communication has been standardized in IEEE 802.11ad for supporting Gigabit WiFi services, mmWave cellular networks may coexist with IEEE 802.11ad systems in the unlicensed spectrum for 5G. In such a scenario, user association may have to be reconsidered in order to make best use of the unlicensed spectrum. To date, this problem has not been investigated yet, and current research efforts focus mainly on user association in integrated LTE-WiFi networks~\cite{Mahindra:2014:PTM:2639108.2639120,Fanqin2015}.} Considering that mmWave networks will also coexist with the diverse HetNets, the following aspects have to be carefully addressed for effective user association design.
\begin{itemize}
  \item \textbf{Large mmWave bandwidth}. Compared to the narrow cellular bandwidths, mmWave cells provide substantially wider  bandwidths, which significantly improves the attainable throughput~\cite{RDaniels_2010}. As such, new user association methods should account for  the effect of system bandwidth.

  \item \textbf{Large array gains}. For a fixed array aperture, the shorter wavelengths of mmWave frequencies enable the mmWave BSs to pack more antennas in the same space, which provides large array gains and therefore increases the received signal power. The simple user association metric based on the minimum-distance rule~\cite{HG2015} may become inefficient, particularly when massive MIMO is applied in  macrocells~\cite{Jungnickel_IEEE_Commag}. The antenna array gains in the mmWave cells will be different from the antenna array gains in the macrocells. As such, new user association methods should also address the effect of large array gains, which is however coupled with very narrow pencil-beams that are hard to update at high velocities.

   \item \textbf{Energy efficiency}. Since mmWave systems use large bandwidths and large antenna arrays, the associated power consumption has to be carefully considered~\cite{SundeepRan_2013}. As such, new energy efficient user association methods are required.

 \end{itemize}

So far, we have only  discussed  the coupled user association based on the downlink. The decoupling access techniques~\cite{Elshaer_H_2014,Smiljkovikj_K2015,Federico_Boccardi2015}, which basically consider the downlink and uplink as separate network entities, may be difficult to be applied in mmWave cellular networks. As mentioned in~\cite{Federico_Boccardi2015}, mmWave beamforming tends to rely on exploiting the uplink/downlink channel reciprocity. An interesting approach suggested in~\cite{Federico_Boccardi2015} is that a user is associated with the mmWave cell BS in the downlink and with a sub 6 GHz macro BS in the uplink.

Given the fact that mmWave solutions are expected to serve as an essential enabling technology in 5G networks, user association in mmWave 5G networks  is a promising research field. In a nutshell, for the potential mmWave component of 5G networks, fundamental research facilitating efficient user association has numerous open facets.

\section{User Association in Energy Harvesting Networks}
One of the main challenges in 5G networks is the improvement of the energy efficiency of radio access networks (RANs) and battery-constrained wireless devices. In the context of prolonging the battery recharge-time and improving the overall energy efficiency of the network, harvesting energy from external energy sources may be viewed as an attractive solution~\cite{HE15}.
In this section, we survey user association in renewable energy powered networks and RF WPT enabled networks, respectively.

Table~\ref{Table_EH} provides a qualitative comparison of the existing user association algorithms proposed for energy harvesting networks.

\begin{table*}[tb]
\newcommand{\tabincell}[2]{\begin{tabular}{@{}#1@{}}#2\end{tabular}}
\renewcommand{\multirowsetup}{\centering}
\scriptsize
\centering
  \caption{Qualitative Comparison of User Association Algorithms for Energy Harvesting Networks.}
\begin{tabular}{|l|r|r|r|r|r|r|r|r|r|r|}
  \hline
  \textbf{Ref.}&\textbf{Algorithm}&\textbf{Topology }&\textbf{Model }& \textbf{Direction}&\textbf{Control }& \textbf{\tabincell{r}{Spectrum\\efficiency}}& \textbf{\tabincell{r}{Energy\\efficiency}}& \textbf{\tabincell{r}{QoS\\provision}}& \textbf{Fairness}&\textbf{\tabincell{r}{Coverage\\probability}}\\
  \hline
   \cite{DHS14} & max-RSS UA &\multirow{2}{0.8cm}{Random spatial} & \multirow{2}{0.8cm}{Stochastic geometry} & DL & Distributed& -& High& Moderate& -& Moderate\\
   \cite{YZ14} & biased UA & & &  DL & Distributed & High & High& - & - &Moderate\\\hline
  \cite{RJ14}  & UA & \multirow{4}{0.8cm}{Grid} & \multirow{4}{1.25cm}{Combinatorial optimization} &  DL & Distributed & High & High & High & High& -\\

  \cite{DL14ICT} & UA &  &  &  DL & Centralized & High & High & High & -& -\\
\cite{ZTLETTER} & UA &  &  &  DL & Distributed & Moderate & High & High & High& -\\

    \cite{JS15} &{\tabincell{r}{UA+power control+\\resource block allocation}}  & & & UL& Centralized & High&High&High & - &- \\\hline
  \cite{DL14PIMRC,DL14GC,DLETT} & UA & \multirow{5}{0.8cm}{Grid} & \multirow{5}{1.25cm}{Combinatorial optimization}& DL & Distributed & Moderate & High& High & - &-\\
  \cite{TH12} & UA &  & & DL & Centralized & - & High& High & - &-\\
  \cite{TH13} & UA+energy allocation &  & & DL & Centralized & - & High& Moderate & - &-\\
  \cite{DLWCNC} & UA+energy allocation &  & & DL & Distributed & - & High& High & - &-\\
  \cite{WB15} & UA+bandwith allocation &  & & DL & {\tabincell{c}{Distributed/\\Centralized}} & - & High& High & - &-\\\hline
  \cite{S_A_H_WPT_2015} &{\tabincell{c}{UA}}  &\multirow{2}{0.8cm}{Random spatial} & \multirow{2}{0.8cm}{Stochastic geometry}&  UL & Distributed & - & High& High & - &Moderate\\
  \cite{S_A_H_Flex_2015} & biased UA  & & &  UL & Distributed & - & High& High & - &High\\
  \hline
\end{tabular}
\label{Table_EH}
\end{table*}

\subsection{User Association in Renewable Energy Powered Networks}
Motivated by environmental concerns and the regulatory pressure for finding ``greener'' solutions~\cite{CAOYUE2015}, network operators have considered the deployment of renewable energy sources, such as solar panels and wind turbines, in order to supplement the conventional power grid in powering BSs. In this scenario, BSs are capable of harvesting energy from the environment and do not require an \emph{always-on} energy source~\cite{LH13}. This is of considerable interest in undeveloped areas, where the power grid is not readily available. Furthermore, it opens up entirely new categories of low cost \emph{drop and play} small cell deployments for replacing the \emph{plug and play} solutions.

\subsubsection{User Association in Solely Renewable Energy Powered Networks}
Due to the randomness of the energy availability in renewable energy sources, integrating energy harvesting capabilities in BSs entails many challenges in terms of the user association algorithm design. The user association decision should be adapted according to the energy and load variations across time and space. The authors of~\cite{DHS14} developed a model for HetNets relying on stochastic geometry, where all BSs were assumed to be solely powered by renewable energy sources. They also provided a fundamental characterization of regimes under which HetNets relying on renewable energy powered BSs have the same performance as the ones benefiting from grid-powered BSs. By relaxing the primal deterministic user association to a fractional user association, the authors of~\cite{RJ14} proposed a user association algorithm for the maximization of the aggregate downlink network utility based on the per-user throughput, where the BSs were solely powered by renewable energy and equipped with realistic finite-capacity batteries. In~\cite{DL14ICT}, adaptive user association was formulated as an optimization problem, which aimed at maximizing the number of supported users and at minimizing the radio resource consumption in HetNets with renewable energy powered BSs. Both optimal offline and online algorithms were developed. Authors of \cite{ZTLETTER} proposed a distributed user association algorithm  for energy consumption and traffic load balancing tradeoffs among heterogeneous base stations in HetNets with renewable energy supply. The deployment of relays having energy harvesting capabilities has also attracted significant attention, since they are able to improve the system capacity and coverage in remote areas which do not have access to the power grid. In this context, Song \emph{et al}.~\cite{YZ14} studied the user association problem targeting the downlink throughput optimization of energy harvesting relay-assisted cellular networks, where BSs were powered by the power grid and relays were powered by the renewable energy. The authors developed a dynamic biased user association algorithm, where the bias was optimized based on the renewable energy arrival rates. In \cite{JS15}, joint user association, resource block allocation, and power control was investigated with the goal of maximizing the uplink network throughput in cellular networks employing energy harvesting relays. The energy-harvesting process was characterized by a time-varying Poisson process. The authors proposed a new metric, referred to as the survival probability, as selection criterion for an energy-harvesting relay.

\subsubsection{User Association in Hybrid Energy Powered Networks}
Although the amount of renewable energy is potentially unlimited, the intermittent nature of the energy from renewable energy sources results in a highly random energy availability at the BS. Thus, BSs powered by hybrid energy sources, which employ a combination of the power grid and renewable energy sources are preferable over those solely powered by renewable energy sources in order to support uninterrupted service~\cite{NDWK13}. The concept of hybrid energy sources has already been adopted by the industry. For instance, Huawei has deployed BSs which draw their energy from both constant energy supplies and renewable energy sources~\cite{HUAWEI}. In the literature, power allocation~\cite{JG2013}, coordinated MIMO~\cite{YC2012}, and sophisticated network planning~\cite{MZ2013} have been studied in the context of cellular networks powered by hybrid energy sources. For the user association designed for such networks, the vital issue is the minimization of the grid energy consumption as well as guaranteeing the user QoS, as detailed in~\cite{DL14GC,DL14PIMRC,DLETT,TH12,WB15,TH13,DLWCNC} and in the references therein. Distributed delay-energy aware user association was proposed for the HetNets operating both with~\cite{DL14GC} and without the assistance of relays~\cite{DL14PIMRC} in order to reduce the grid power consumption by maximizing the exploitation of ``green'' power harvested from renewable energy sources, as well as to enhance the QoS by minimizing the average traffic delay. Extending the solutions of~\cite{DL14PIMRC}, the authors of~\cite{DLETT} addressed the backhaul constraint and the uplink-downlink asymmetry in the context of designing the user association algorithm for HetNets relying on hybrid energy sources. In~\cite{TH12}, an intelligent cell breathing method was proposed for minimizing the maximal harvested energy depletion rate of BSs. In~\cite{WB15}, a constrained total energy cost minimization problem was formulated, which was then solved with the aid of energy efficient user association and bandwidth allocation algorithms. Multi-stage harvested energy allocation and multi-BS traffic load balancing algorithms were designed for energy optimization in~\cite{TH13}. In~\cite{DLWCNC}, two-dimensional optimization of user association in the spatial dimension and harvested energy allocation in the time dimension was developed for minimizing both the total and the maximum grid energy consumption, while guaranteeing a certain QoS. Extending from~\cite{DLWCNC}, online algorithms for two-dimensional optimization were further developed in~\cite{DTTCOM}.

\subsubsection{User Association in Energy Cooperation Enabled Networks}
Because of the spatial and temporal dynamics of the renewable energy, BSs may suffer from geographical variability in terms of their harvested renewable energy. Fortunately, the recent development of the smart grid facilitates two-way information and energy flows between the grid and the BSs of cellular networks~\cite{FX12}, which makes \emph{energy cooperation} between BSs possible. Energy cooperation between BSs allows the BSs that have excessive harvested renewable energy to compensate for others that have a deficit due to either higher traffic load or lower generation rate of renewable energy, thereby substantially improving the renewable energy utilization and decreasing the grid energy consumption. The concept of \emph{energy cooperation} has inspired significant research efforts in recent years, see~\cite{YKC14,YG14,XJ14} and references therein. The optimal energy cooperation policy conceived for minimization of the system grid energy consumption was disseminated in~\cite{YKC14}. Joint energy and spectrum cooperation invoked for the minimization of the energy cooperation cost was developed in~\cite{YG14}. {In~\cite{XJ14}, the weighted sum rate of all users was maximized by joint power allocation and energy cooperation optimization in CoMP aided networks.}

\begin{figure} [tb]
\vspace{-0.cm}
    \centerline{\includegraphics[width=0.45\textwidth,height=0.3\textwidth]{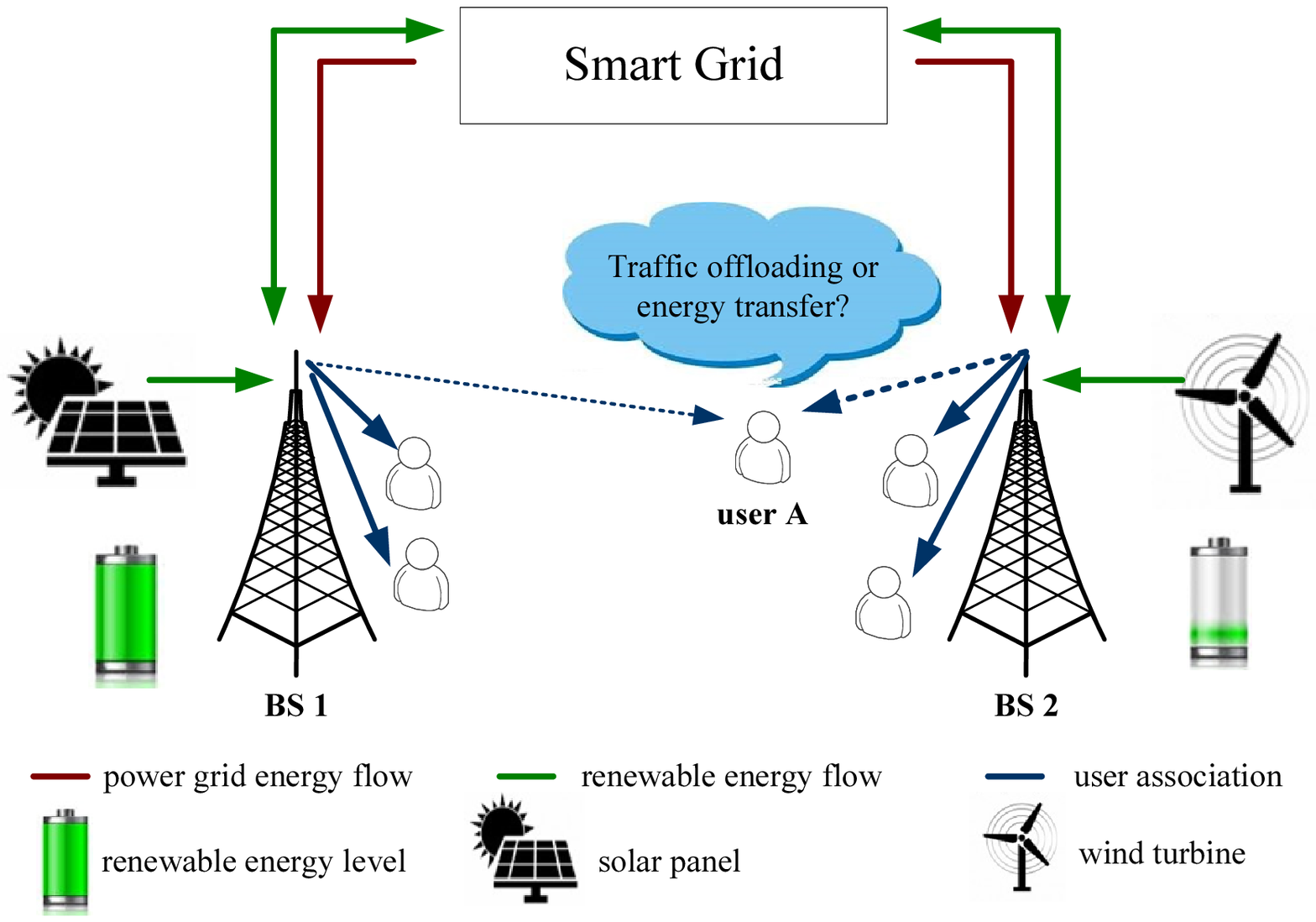}}
    \caption{The tradeoff between traffic offloading and energy transfer in energy cooperation enabled networks.}
    \label{fig:1}
    \vspace{-0.7cm}
\end{figure}

To the best of our knowledge, research results on user association in energy cooperation enabled networks are not available yet. User association in energy cooperation enabled networks introduces tradeoffs between traffic offloading and energy transfer, which is a challenging research topic. As shown in Fig.~\ref{fig:1}, the stored renewable energy level of BS 2 is much lower than that of BS 1, due to the low renewable energy generation of BS 2. Additionally, more users are located in the vicinity of BS 2. If both BS 1 and BS 2 use the same fixed transmit power, BS 2 will have to consume more renewable energy transferred from BS 1 or more energy from the power grid, in order to serve all the users in its vicinity. However, there will be some renewable energy loss during the energy transfer from BS 1 to BS 2. Alternatively, some users near BS 2 may choose to associate with the far-away BS 1 having more harvested renewable energy. For instance, user A in Fig.~\ref{fig:1} may be associated with BS 1, which consequently avoids the renewable energy loss owing to energy transfer. Nonetheless, we observe that user A may suffer from more signal strength degradation, when it is offloaded to BS 1, since the distance between user A and BS 1 is larger than that between user A and BS 2. As such, it is crucial to strike a tradeoff between the signal strength degradation caused by traffic offloading and the renewable energy loss caused by energy transfer with the aid of user association optimization in energy cooperation enabled networks.

Table~\ref{table:3} summarizes the challenges of user association designed for renewable energy powered networks in different scenarios.

\begin{table*}
\newcommand{\tabincell}[2]{\begin{tabular}{@{}#1@{}}#2\end{tabular}}
  \centering
   \caption{Summary of User Association for Renewable Energy Powered Networks}
  \begin{tabular}{|l|r|r|r|}\hline
\textbf{Classification} &\textbf{Scenario} & \textbf{Challenges}& \textbf{Ref.} \\ \hline
\tabincell{l}{User association in \\solely renewable energy\\ powered networks} & \tabincell{r}{BSs in networks are solely powered by \\renewable energy from environment,\\such as solar energy, wind energy.} & \tabincell{r}{User association decision should be adapted\\ according to the renewable energy and load variations \\across time and space. QoS provision may be \\deteriorated by insufficient renewable energy.} & \cite{DHS14,RJ14,DL14ICT,ZTLETTER,YZ14,JS15}\\ \hline
\tabincell{l}{User association in\\ hybrid energy \\powered networks} & \tabincell{r}{BSs in networks are powered by\\ a combination of power grid \\and renewable energy sources.} & \tabincell{r}{User association decision should maximize the \\utilization of renewable energy, minimize the grid \\energy consumption and guarantee the QoS provision.}& ~\cite{DL14GC,DL14PIMRC,DLETT,TH12,WB15,TH13,DLWCNC} \\ \hline
\tabincell{l}{User association in\\ energy cooperation \\enabled networks} & \tabincell{r}{BSs with excess harvested renewable\\ energy can aid other BSs with\\ energy shortage via renewable energy transfer.} & \tabincell{r}{User association decision is crucial for the tradeoff \\between signal strength degradation caused by traffic\\ offloading and energy loss in energy transfer.}& --\\ \hline
\end{tabular}
\label{table:3}
\end{table*}

\subsection{User Association in RF WPT Enabled Networks}
Traditional energy harvesting sources, such as solar, wind, hydroelectric, etc., depend on locations and environments. RF WPT is an alternative approach conceived for prolonging the lifetime of mobile devices~\cite{CheDZ14,Yuanwei2014_globcom,huangkaibin_mag_2015,S_A_H_WPT_2015,H_T_WPT_2015,Lifeng_wang2015_globcom}. The advantages of WPT are at least two-fold: 1) Unlike the traditional energy harvesting, it is independent of the environment's conditions, and can be applied anywhere; 2) It is flexible and can be scheduled at any time. Additionally, the potentially harmful interference received by the energy harvester actually becomes a precious energy source.

User association in cellular networks relying on ambient RF energy harvesting has been studied in~\cite{S_A_H_WPT_2015,S_A_H_Flex_2015}. In~\cite{S_A_H_WPT_2015}, an analytical approach considering K-tier uplink cellular networks with RF energy harvesting was presented, where mobile users relied only on the energy harvested from ambient RF energy sources for powering up their devices for uplink transmissions. The ``harvest-then-transmit'' strategy was adopted by the users. This contribution  performed uplink user association based on the best average channel gain, i.e., the lowest path-loss. The authors of~\cite{S_A_H_Flex_2015} further examined RF energy harvesting in the context of HetNets in conjunction with flexible uplink user association. In the flexible user association, users were not necessarily associated with their nearest BS, because a different bias factor was added to each network tier.

As pointed out in~\cite{huangkaibin_mag_2015}, the RF energy scavenging considered in \cite{S_A_H_WPT_2015,S_A_H_Flex_2015} is only sufficient for powering small sensors, and dedicated power beacons have to be employed for  powering larger devices. More particularly, in the HetNets associated with dense small cells, the distance between the users and the serving BSs is typically shorter, which suggests that the serving BS can act as a dedicated RF energy source to power its user, similar to power beacons. Hence, there are two types of user association designs for WPT in cellular networks: 1) Downlink based user association for maximizing the harvested energy, which increases the user's transmit power; 2) Uplink based user association for minimizing the uplink path-loss, which increases the received signal power at the serving BS \cite{S_A_H_WPT_2015,S_A_H_Flex_2015}. As such, the uplink-downlink decoupling access studied in~\cite{Elshaer_H_2014,Smiljkovikj_K2015,Federico_Boccardi2015} is a promising approach to achieve both  maximum downlink harvested energy and minimum uplink path-loss. However, these research topics have not been well investigated at the time of writing, and hence they require further exploration.
\subsection{Summary and Discussions}
The deployment of renewable energy sources to supplement the conventional power grid for powering BSs indisputably underpins the trend of green communication. However, the intermittent nature of renewable energy sources requires a rethinking of the traditional user association rules designed for conventional cellular networks relying on constant grid power supply. The existing research contributions regarding user association for renewable energy powered networks aim for maximizing the exploitation of renewable energy, while maintaining the QoS guarantees. On the other hand, the smart grid, as one of the use cases envisioned for 5G networks~\cite{Nokia}, has paved the way for energy cooperation in networks. Energy cooperation between BSs allows the BSs that have excessive harvested renewable energy to assist other BSs that have an energy deficit via renewable energy transfer. To the best of our knowledge, user association in energy cooperation enabled networks is still an open field, and is expected to become a rewarding research area.

Additionally, the existing research contributions on energy harvesting networks investigate user association in networks relying on either renewable energy powered BSs or RF energy harvesting assisted users. User association design in networks combining renewable energy powered BSs and RF energy harvesting assisted users is still untouched and is expected to be a promising approach for 5G networks. In such a network scenario, BSs are capable of harvesting renewable energy from the environment, such as solar power and wind power, while users are powered by RF energy harvesting, thereby having the promise of dramatically reducing the energy consumption of BSs as well as prolonging the battery recharge time. Nevertheless, in this context, both the renewable energy harvested by the BSs and the RF energy harvested by the users will simultaneously play a crucial role in determining the user association, where the user association algorithm should be carefully redesigned for adequate QoS provision and energy consumption reduction.

\section{{User Association in Networks employing other Technologies for 5G}}
{In the previous sections, user association was investigated by emphasizing the impact of key 5G techniques including HetNets, massive MIMO, mmWave, and energy harvesting. Needless to say that there are other important technologies for 5G. Their impact on user association will be discussed in this section.}
\subsection{Self-Organizing Networks}
{In the SONs with the ability of self-configuration, self-optimization, and self-healing, the amount of required  manual work is minimized in order to reduce the OPEX~\cite{HonglinHu_2010}. The specific requirements and use cases for SONs have been summarized and discussed in  standards and industry organizations, such as 3GPP and the next-generation mobile networks alliance~\cite{Mugen_Peng_COMST}. In SONs, there are multiple use cases for network optimization such as capacity and coverage optimization (CCO) as well as mobility load balancing (MLB)~\cite{Fehske_AJ_2013}. In \cite{Fehske_AJ_2013}, $\alpha$-optimal user association was adopted and an algorithm  for optimizing both the user association and the antenna-tilt setting was introduced for  the CCO and MLB SON use cases. In \cite{Lobinger_A_2011},  the  coordination of RSS-based handover and load balancing of SON algorithms was examined in the context of LTE networks, which aimed at combining the strengths of both algorithms.}

\subsection{Device-to-Device Communication}
{D2D communication supports  direct transmission based proximity services between devices without the assistance of the BS or the core network, in an effort to improve the spectrum and energy efficiency~\cite{Asadi_A_2014,Yuanwei_Liu_lifeng_2015}. D2D communication can be operated in inband D2D mode (similar to the cognitive radio networks) or outband D2D mode.  However, one of the key characteristics of D2D is the involvement of the cellular network in the control plane~\cite{Asadi_A_2014}. In \cite{ElSawy_2014_TCOM}, flexible mode selection with truncated channel inversion power control was analyzed in  underlay D2D cellular networks, where users chose the D2D mode or were connected with  BSs based on the uplink quality. In \cite{Vlachos_2015},  D2D link cell association was studied and an optimization approach was proposed  for reducing signalling load and latency in network control based D2D links. In \cite{Semiari_O_2015}, the D2D link was established based on the social influence of the D2D transmitter that owns the popular content of common interest.}

\subsection{Cloud Radio Access Network}
{As a new mobile network architecture consisting of RRHs and BBUs, C-RAN is capable of efficiently dealing with large
scale control/data processing. The rationale behind this approach is that
baseband processing is centralized and coordinated among
sites in the centralized BBU pool, which reduces both the CAPEX
and OPEX~\cite{checko_2015}. In addition, the C-RAN mitigates the
inter-RRH interference by using efficient interference management
techniques such as CoMP. In \cite{Shixin_Luo_TWC_2015}, joint downlink
and uplink user association and beamforming design for
C-RANs was proposed for minimizing the power consumption
under downlink and uplink QoS constraints. In \cite{Heli_Zhang_2015}, the user
association optimization problem was formulated for minimizing
the network's latency, and a three-phase search algorithm was
introduced for solving it. In \cite{Zaidi_2015_COMML}, user-centric association was
adopted for maximizing the downlink received signal-to-noise
ratio (SNR) in the C-RAN with the aid of stochastic geometry, where both the
coverage probability and downlink throughput were analyzed.}

\subsection{Full-Duplex Communication}
{Full-duplex communication is capable of potentially doubling the
spectrum efficiency by allowing simultaneous downlink and
uplink transmission within the same frequency band~\cite{Kim_D_2015}. However, self-interference (SI) suppression becomes critical
in full-duplex systems, since it will seriously deteriorate the
reception quality. Many SI mitigation methods have
been studied. More particularly, in \cite{Zhongshan_2015,Kim_D_2015}, the authors comprehensively
investigated  various SI mitigation methods
by considering both passive and active techniques. In \cite{Goyal_S_2015}, a
hybrid scheduling scheme was presented, where the users were
scheduled in half-duplex or full-duplex mode to avoid imposing
excessive interference on the network. In \cite{Sekander_2015}, the feasibility
of decoupled uplink-downlink user association in full-duplex two-tier cellular networks
was investigated, and a matching game was formulated for
maximizing the total uplink and downlink throughput.}

\subsection{Summary and Discussions}
{Although the aforementioned techniques can effectively enhance spectrum efficiency and energy efficiency with lower CAPEX and OPEX, their features pose substantial challenges to user association design. For example, in self-organizing cellular networks, user association has to be adapted to the specific requirements imposed by SON. When full-duplex communication is employed in cellular D2D networks, three cases may be distinguished, namely 1) the D2D link is half-duplex, and  the normal link via the BS is   full-duplex; 2) the D2D link is full-duplex, and  normal link via the BS is  half-duplex; and 3) both the D2D link and the normal link via the BS are full-duplex. For each case, user association has to be carefully redesigned to reduce the interference. In the C-RAN, considering  that inter-RRH interference can be efficiently  mitigated via cooperation among RRHs, user association algorithms designed for interference coordination are obsolete. In addition, user-centric association may become preferable to the current BS-centric one as a large number  of RRHs will be deployed in the C-RANs~\cite{Zaidi_2015_COMML}.}

{Current research efforts have provided a good understanding of the aforementioned technologies. Nevertheless, the number of studies of user association mechanisms for  networks employing SONs, D2D, C-RAN and full-duplex communication is limited.  Hence, more research into this direction is needed in the future.}

\section{Summary and Conclusions}
{The pertinent user association algorithms designed for HetNets,
massive MIMO networks, mmWave scenarios and energy harvesting networks
have been surveyed, which constitute four of the most salient enabling
technologies envisioned for future 5G networks. In order to
systematically survey the existing user association algorithms, we
have presented a related taxonomy. Within each of the networks
considered, we have highlighted the inherent features of the
corresponding 5G enabling technology, which have a substantial impact
on the user association decision, and then categorized the
state-of-the-art user association algorithms.  However, given the
intricate and perpetually evolving 5G network conditions, the related
research relying on sophisticated machine learning techniques is still
in its infancy. Hence, a range of challenging open issues regarding
user association in 5G networks have also been summarized in this
paper. Indeed, user association has to be investigated in more depth
as a community-effort in order to better accommodate the inherent
features of 5G enabling technologies, so as to realize the full
potential of 5G networks.}

{When designing and optimizing a wireless system, the most influential
factor in predetermining the overall performance of the system is the
specific choice of the metric to be optimized.  For example, when we
aim for an increased bandwidth efficiency, we opt for high-throughput
modulation schemes, which are however not energy-efficient. The
opposite is true for the family of power-efficient, but low-throughput
$m$-ary orthogonal modulation, which operates at low SNRs.}

{In this spirit, in order to make our discussions well-balanced, we
have used five classic metrics throughout this treatise, namely the
outage/coverage probability, bandwidth efficiency, energy efficiency,
QoS and fairness. It is feasible to specifically choose the user
association metrics in order to satisfy the prevalent user
requirements. For example, when the users request high-definition
video streaming services, the bandwidth efficiency may be the
preferred metric to be optimized, whilst compromizing the energy
efficiency and vice versa. The choice of this metric is much more
crucial than the choice of the optimization algorithms and tools
invoked for optimizing it!}

{Bearing in mind the fact that 5G is expected to become the fusion of
heterogeneous wireless technologies, the user association problem
should be tackled by giving careful cognizance to the specific 5G
scenario encountered, in order to satisfy the tight specifications of the
enabling techniques considered in this survey. For example, in the
areas where massive MIMO-aided macrocells and mmWave small cells are
employed, user association should be designed based on the
detailed recommendations of Section V-C.}

{In the context of energy-efficient harvesting-aided networks the BSs
may be powered by renewable energy sources. As such, the energy
consumption constraints play a crucial role in influencing the user
association design, as mentioned in Section VI.  When the handsets are
recharged with the aid of RF WPT, the BSs may act as RF energy
sources. For example, in such scenarios the user association
techniques may be specifically designed for receiving the maximum
possible amount of RF energy.  Naturally, a diverse variety of other
compelling user association designs are possible for the sake of
enhancing the attainable performance by considering the basic design
principles outlined above.}

\end{document}